\documentstyle[epsf]{l-aa} 
\begin{document}

\newcommand{\Kkms}{\,{\rm K\,km/s} }
\newcommand{\WHI}{\,{W_{\rm HI}} } 

\newcommand{\mic}{\,{\rm \mu m} } 
\newcommand{\pac}{\,{\rm pc} } 

\newcommand{\Wmsr}{\,{\rm W/m^2/sr} }
\newcommand{\Wcmsr}{\,{\rm W/cm^2/sr} }
\newcommand{\Wcm}{\,{\rm W/cm^2} }
\newcommand{\cmun}{\,{\rm cm^{-1}} }
\newcommand{\Mj}{\,{\rm MJy/sr} }

\title{Implications of the cosmic infrared background for light production and the star formation history in the Universe}

\thesaurus{02(12.03.3
           12.04.2, 13.09.1)}

    \author{R. Gispert  
          \inst{\dag}\thanks{Richard Gispert died in the final stages of the preparation
of this paper in August 1999.} \and 
          G. Lagache \and  
          J.L. Puget }

 \offprints{G. Lagache, lagache@ias.fr} 

\institute{Institut d'Astrophysique Spatiale, B\^at.  121, 
Universit\'e Paris XI, F-91405 Orsay Cedex}

 \date{}  
 
   \maketitle
\markboth{Implications of the cosmic infrared background for the star formation history in the Universe}{}

\begin{abstract}
The Cosmic Background due to the integrated radiation from galaxies over
the whole life of the
Universe is reviewed. We find that this background is well constrained by
measurements. The
total power in the background is in the range 60-93 nWm$^{-2}$sr$^{-1}$. The data
show the existence of a minimum separating the direct stellar radiation from the infrared
part due to radiation reemitted by dust. This reemitted dust
radiation is about 1-2.6 time the background power in the optical/near-IR
thus much larger than the same ratio measured locally ($30\%$). The far-infrared
and submillimeter
background is likely to be dominated by redshifted infrared galaxies. The long wavelength spectrum of 
the background being significantly flatter than the spectrum of these
galaxies it strongly constrains the far-infrared radiation production rate history which must increase by a factor larger
than 10 between the present time and a redshift 1 and then stays
rather constant at higher redshift, contrary to the ultraviolet radiation production rate which
decreases rapidly. 

Several models of galaxy evolution have been proposed to explain the submillimeter background. In this paper 
we do not propose a new model; we systematically explore the allowed range of evolution histories allowed by the data.
If infrared galaxies are mostly powered by starbursts as indicated by recent observations, this infrared production 
history reflects the history of star formation in the Universe.
\end{abstract}
\section{Introduction}

The history of star formation in the Universe is one of the key function in
physical cosmology. It is
closely linked to galaxy formation and evolution and controls the second
most important
contribution to the cosmic electromagnetic background after the Cosmic
Microwave
Background (CMB) generated at the time of recombination at a redshift
around 1000.  It has
been pointed out many times in the past 30 years, that measurements of the
cosmic background
radiated by all galaxies over the history of the Universe would be
extremely valuable for
physical cosmology. It would strongly constrain models for galaxy formation
and evolution
(see for example Partridge \& Peebles, 1967). This background is expected
to be composed of three main components:\\
- The stellar radiation in galaxies concentrated in the ultra-violet and
visible with a redshifted component in the near InfraRed (IR) \\
- A fraction of the stellar radiation absorbed by dust either in the
galaxies or in the intergalactic medium\\
- The radiation from active galactic nuclei (a fraction of which is also absorbed
by dust and reradiated in the far-IR).\\
The energy in the first two components is derived from nucleosynthesis in
stars, the last one
probably derived from gravitational energy of accreted matter onto massive
black holes.
In the last two years the cosmic background at visible, IR and
submillimeter (submm) wavelengths has been finally constrained by very deep
source counts and upper limits on the diffuse isotropic emission at shorter
wavelengths, and measured in the submm range.
We review the observational situation in Sect. \ref{observ_situation}.
In Sect. \ref{formalism}, we define the formalism of the determination
of the IR radiation production rate history. 
When the spectrum of the sources dominating the
background is strongly peaked around a wavelength of 80 $\mu$m, 
the radiation production rate as a function of redshift
can be directly inferred from the spectrum of the cosmic background.
In Sect. \ref{source_spectra}, the sources of Cosmic Far-IR BackgRound (CFIRB)
and their spectra are rewiewed. In the likely hypothesis of this background being dominated by IR 
galaxies (either starburst galaxies or dust enshrouded AGNs), 
we derive the IR radiation production rate as a function of redshift 
in the Universe (Sect. \ref{inversion}). We compare it with other means to measure the star formation rate
in Sect. \ref{sfr}.\\

The study presented here is different from the many studies of this type
based on a model for Spectral Energy Distribution
(SED) and evolution of IR galaxies which, after
some adjustements, account for the Cosmic Background. Considering
the diversity of SED which can be used and the observational
uncertainties it is clear that a range of histories of the
IR galaxies is possible. The goal of this paper and of the method
presented here is to explore systematically the range of possible
solutions by a careful analysis of possible SEDs taking properly
into account the observational uncertainties. The method developed
in this paper is the only one so far to answer this question.

\section{\label{observ_situation} The Cosmic Background Radiation: observations}

\begin{figure*}  
\vspace{-0.6cm}
\epsfbox{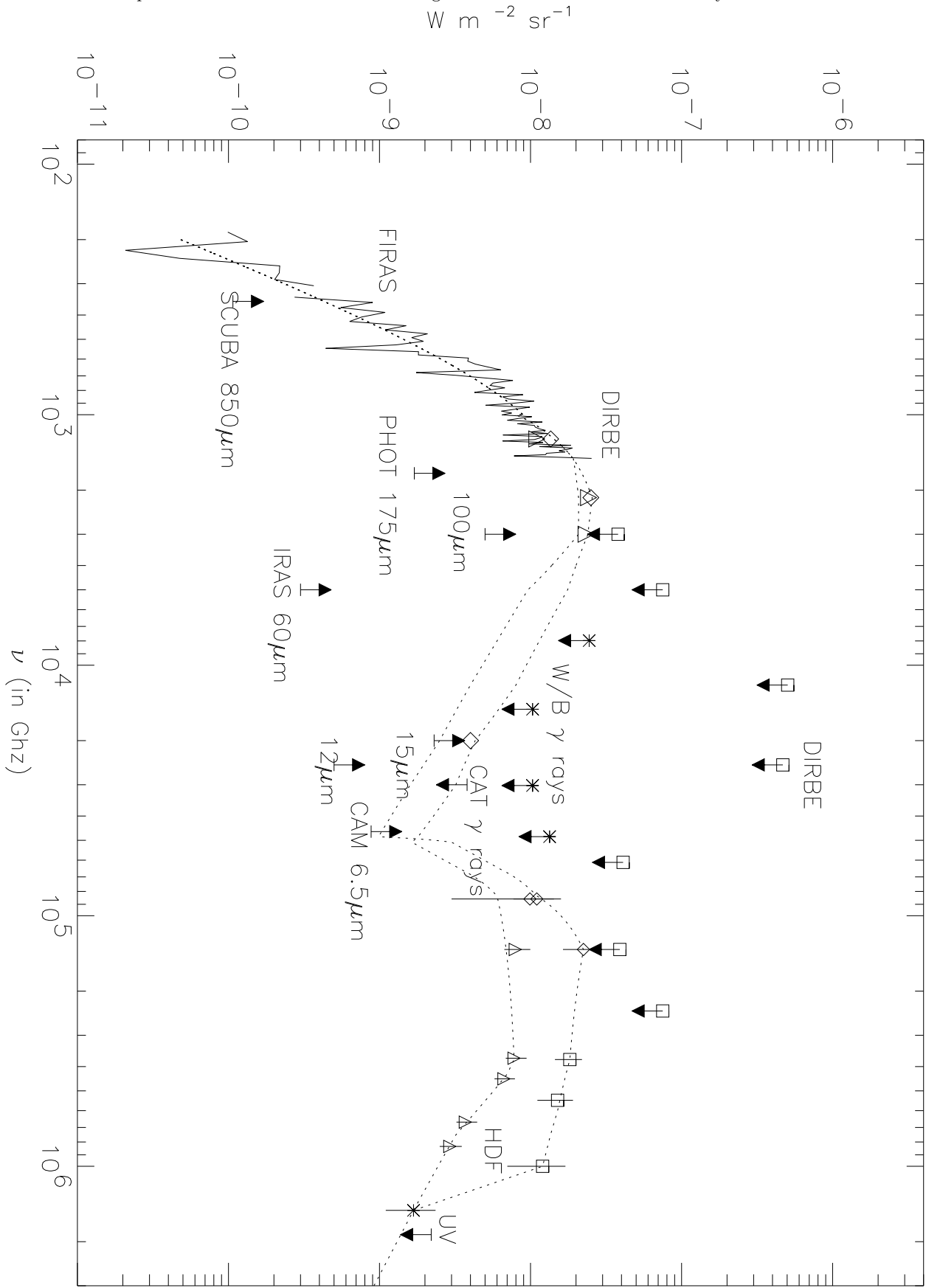}
\vspace{-1.cm}
\caption{ \label{back_all} Cosmic Background from the UV to the
millimeter wavelength. In the UV domain, the upper limit is from Martin et 
al. (1991) and the value from Armand et al. (1994); the optical and near-IR 
points are from Pozzetti et al., 1998 (triangles) and Bernstein et al., in preparation (squares); 
The 3.5 and 2.2 $\mu$m points are from Dwek \& Arendt (1998) and Gorjian et al. (2000); Squared upper limits
are from Hauser et al. (1998) and crossed upper limits from Biller et al. (1998);
the upper limit ``CAT'' is from Barrau (1998) and Barrau et al., in preparation; The 6.5
(D\'esert, private communication), 12 (Clements et al. 1999) and 15 $\mic$ (Elbaz et al 1999)
lower limits come from ISOCAM number counts; the value at 15$\mic$ 
($\diamond$) is an extrapolation of the counts using the Guiderdoni et al. 
(1998) model. At longer wavelength, we have the 100, 140 and 240 $\mic$ 
Lagache et al. 2000 ($\triangle$) and Hauser et al. 1998 ($\diamond$)
DIRBE values, lower limit from Dwek et al. (1998), upper limits from number 
counts at 60 (Lonsdale et al. 1990), 
175 (Puget et al. 1999) and 850 $\mic$ 
(Barger et al. 1999). Dotted lines are an attempt 
to draw continuous lines compatible with all available data which then
can be used for estimating the energy contribution of a given wavelength range.} 
\end{figure*}

We summarise here the observational situation of the cosmic background which is the main observation used in this study. 
The values of the cosmic background at all wavelengths are discussed
in Appendix 1 and shown on Fig. 1.
We can compute the energy contained in the background above
and below 6 $\mu$m using the best fits presented
on Fig. \ref{back_all}. We obtain:\\
- E($<$6 $\mu$m)=E(opt)= 2-4.1 10$^{-8}$ W m$^{-2}$ sr$^{-1}$\\
- E($>$6 $\mu$m)=E(FIR)= 4.-5.2 10$^{-8}$ W m$^{-2}$ sr$^{-1}$\\
We know that for local galaxies the ratio E(FIR)/E(opt)
is about 0.3 (Soifer \& Neugeubauer, 1991). 
We measure here a higher ratio which is about 1-2.6.
This higher ratio results from two combined
effects. First, the redshift effect will bring the light from
optical to IR and thus will give more energy in the IR
than for local galaxies. Nevertheless this effect cannot account for an IR to optical ratio 
much larger on average than the value observed today. This higher ratio can only be
explained by a change of properties of galaxies that are
making the background in the optical and the IR.
A very simple comparison supports this idea: the
energy in the backgound at 15 $\mu$m and in the optical
domain is nearly the same. But the background at 15 $\mu$m in  the Hubble Deep Field 
is made by a very small number of objects as compared to the number of sources which contribute to
the bulk of the energy observed in the optical (Aussel et al. 1999). This means that the background
in the IR is probably dominated by few objects with high IR luminosities.
Thus these objects are not the standard IR counterparts
of normal spiral and irregular galaxies.

\section{\label{formalism} Formalism of the determination of the radiation production rate 
history in the Universe}

For a Universe following a Robertson-Walker metric with total density
$\Omega_0$, and a cosmological constant $\Omega_{\Lambda}$,  the 
differential contribution dE to the background energy density 
(monochromatic or integrated over frequency, per unit volume)
generated around redshift z during time dt is:
\begin{equation}
dE= \frac{\varphi (z)}{(1+z) }dt
\end{equation}
where $\varphi$ is the comoving energy production rate per
unit volume at redshift z (note that here $\varphi$ is not the spectral
energy density).

The backgound spectral intensity (brightness per solid angle and frequency
interval) I$_{\nu}$ is such that:
\begin{equation}
\label{very_simple}
\nu I_{\nu}= \frac{c}{4 \pi} \varphi (z) \left| \frac{dt}{dz} \right| _{1+z=\frac{\nu^{'}}{\nu}}
\end{equation}
where we assume a monochromatic spectrum for the sources
radiating at frequency $\nu^{'}$=$\nu$ (1+z) where $\nu$ is the observed
frequency.
In that case ($\delta$ function spectrum for the emitting objects),
$\varphi (z)$ can be simply deduced from
the background spectrum $\nu I_{\nu}$ (see Appendix 2).\\

The relations given above can be easily generalised to the case where the
sources radiate through a broad spectrum:
\begin{equation}
\label{eq_principale}
\nu I_{\nu}= \frac{c \nu}{4 \pi} \int_{\nu^{'}=\nu}^{\infty} N_{z} L_{\nu^{'}} \left| \frac{dt}{dz}\right| d\nu^{'}
\end{equation}
with $N_{z}$ the number of sources per Mpc$^3$ and $L_{\nu^{'}}$ the luminosity
of the galaxies (in W/Hz).\\
$L_{\nu^{'}}$ is the average over a large volume of SED of 
luminous IR galaxies which have been shown to dominate the 15 $\mu$m
background at $z \simeq 0.7$ (Aussel et al 1999).
The range of SED for IR galaxies are discussed in 
Sect. \ref{source_spectra}. The important result of this discussion is that all the far-IR SEDs 
of luminous IR galaxies are not very dependent of the energy source (starburst or dust enshrouded AGN) 
and varies slowly with the luminosity (Maffei, 1994; Guiderdoni 1998). \\
The unknown quantity in Eq. (\ref{eq_principale}) becomes $N_{z}$.
The full range of allowed functions for
the number density of sources as a function of z is derived using Monte Carlo simulations
for a set of IR galaxy spectra which covers the range of possible ones and
for different cosmological models.
The number density multiplied by the total luminosity of each model galaxy
gives the IR luminosity density $\varphi$ in function of z:
\begin{equation}
\label{GPL_k}
\varphi(z)= N_{z} \int_{\nu_0}  L_{\nu_0} d\nu_0 
\end{equation}
$\varphi(z)$ depends on the cosmological model via the $\frac{dt}{dz}${\footnote{We can note that $\varphi(z) \times \frac{dt}{dz}$ is independent of the cosmological model. We nevertheless choose to use $\varphi(z)$ rather $\varphi(z) \times \frac{dt}{dz}$ as $\varphi(z)$ correponds at z=0 to a measured value.}}.

\begin{center}
\begin{figure}  
\epsfxsize=9.cm
\epsfysize=7.cm
\epsfbox{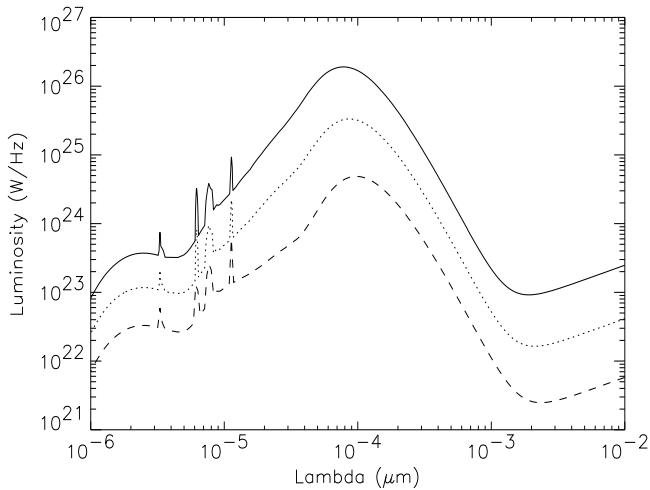}
\caption{ \label{fig_spec_gal}Typical spectra of starburst galaxies (Maffei, 1994)
for different luminosities
(continuous line: 3 10$^{12}$ L$_{\odot}$, dotted line: 5 10$^{11}$ L$_{\odot}$,
dashed line: 7 10$^{10}$ L$_{\odot}$) with a spectral dust emission index in
the far-IR of 2.}
\end{figure}
\end{center}

\begin{table*}
\tiny
\caption{Co-moving luminosity density $\varphi$ derived from the CFIRB for the three galaxy model luminosities,
the four dust spectral indexes and the three cosmologies (defined by $\Omega_{\Lambda}$ and $\Omega_{0}$).
Numbers on the rigth side of each row correspond to the miminal and maximal
value of $\varphi$ compatible with the CFIRB.
The Hubble constant is the same for all cases (h=0.65). Given in column 10 is the $\chi^2$ of
the CFIRB ``fit'' derived from the luminosity density compared to the FIRAS and DIRBE measured values.}
\label{tbl-1}
\begin{center} 
\begin{tabular}{|c||c|c|c|c|c|c|c|c|c|} \hline
Log(L), $\alpha$, $\Omega_{\Lambda}$, $\Omega_{0}$ & \multicolumn{7}{|c|}{Luminosity density $\varphi$ (L$_{\odot}$/Mpc$^3$)} & $\chi^2$ \\ 
 & \multicolumn{1}{c}{z=0.4} & \multicolumn{1}{c}{z=0.8}  & \multicolumn{1}{c}{z=1.2} & \multicolumn{1}{c}{z=1.9} & \multicolumn{1}{c}{z=2.7} & \multicolumn{1}{c}{z=4.8} & \multicolumn{1}{c|}{z=8} &  \\ \hline
12.5, 2.0, 0.0, 0.3 &8.94 $10{^8}_{8.28\hspace{0.07cm} 10{^ 6}}^{8.28\hspace{0.07cm} 10{^ 9}}$  &1.15 $10{^9}_{6.75\hspace{0.07cm} 10{^ 8}} ^{1.45\hspace{0.07cm} 10{^ 9}}$  &1.53 $10{^9}_{1.04\hspace{0.07cm} 10{^ 9}} ^{2.08\hspace{0.07cm} 10{^ 9}}$  &1.26 $10{^9}_{8.61\hspace{0.07cm} 10{^ 8}}^{1.85\hspace{0.07cm} 10{^ 9}}$  &1.43 $10{^9}_{7.76\hspace{0.07cm} 10{^ 8}} ^{1.95\hspace{0.07cm} 10{^ 9}}$  &6.68 $10{^8}_{6.68\hspace{0.07cm} 10{^ 6}} ^{1.23\hspace{0.07cm} 10{^ 9}}$  &3.82 $10{^8}_{4.13\hspace{0.07cm} 10{^ 6}} ^{7.07\hspace{0.07cm} 10{^ 8}}$  &0.054 \\ \hline
12.5, 2.0, 0.7, 0.3 &6.36 $10{^8}_{1.27\hspace{0.07cm} 10{^7}} ^{1.01 \hspace{0.07cm} 10^{10}}$  &1.54 $10{^9}_{9.74\hspace{0.07cm} 10{^ 8}} ^{1.94\hspace{0.07cm} 10{^ 9}}$  &1.91 $10{^9}_{1.40\hspace{0.07cm} 10{^ 9}} ^{2.59\hspace{0.07cm} 10{^ 9}}$  &1.86 $10{^9}_{1.27\hspace{0.07cm} 10{^ 9}}^{2.53\hspace{0.07cm} 10{^ 9}}$  &2.01 $10{^9}_{1.18\hspace{0.07cm} 10{^ 9}} ^{2.96\hspace{0.07cm} 10{^ 9}}$  &1.14 $10{^9}_{3.34\hspace{0.07cm} 10{^7}} ^{2.11\hspace{0.07cm} 10{^ 9}}$  &5.64 $10{^8}_{4.83\hspace{0.07cm} 10{^ 6}} ^{1.21\hspace{0.07cm} 10{^ 9}}$  &0.053 \\ \hline
12.5, 2.0, 0.0, 1.0 &3.02 $10{^8}_{1.77\hspace{0.07cm} 10{^ 6}} ^{8.20\hspace{0.07cm} 10{^ 9}}$  &1.36 $10{^9}_{8.60\hspace{0.07cm} 10{^ 8}} ^{1.85\hspace{0.07cm} 10{^ 9}}$  &1.96 $10{^9}_{1.24\hspace{0.07cm} 10{^ 9}} ^{2.67\hspace{0.07cm} 10{^ 9}}$  &1.73 $10{^9}_{1.18\hspace{0.07cm} 10{^ 9}}^{2.54\hspace{0.07cm} 10{^ 9}}$  &2.06 $10{^9}_{1.12\hspace{0.07cm} 10{^ 9}} ^{2.81\hspace{0.07cm} 10{^ 9}}$  &1.04 $10{^9}_{1.04\hspace{0.07cm} 10{^7}} ^{2.07\hspace{0.07cm} 10{^ 9}}$  &6.75 $10{^8}_{6.75\hspace{0.07cm} 10{^ 6}} ^{1.25\hspace{0.07cm} 10{^ 9}}$  &0.053 \\ \hline
11.7, 2.0, 0.0, 0.3 &1.50 $10{^9}_{8.77\hspace{0.07cm} 10{^ 6}} ^{7.52\hspace{0.07cm} 10{^ 9}}$  &1.37 $10{^9}_{8.64\hspace{0.07cm} 10{^ 8}} ^{1.72\hspace{0.07cm} 10{^ 9}}$  &1.49 $10{^9}_{1.01\hspace{0.07cm} 10{^ 9}} ^{2.19\hspace{0.07cm} 10{^ 9}}$  &1.17 $10{^9}_{7.95\hspace{0.07cm} 10{^ 8}}^{1.71\hspace{0.07cm} 10{^ 9}}$  &1.25 $10{^9}_{6.28\hspace{0.07cm} 10{^ 8}} ^{1.70\hspace{0.07cm} 10{^ 9}}$  &4.15 $10{^8}_{3.84\hspace{0.07cm} 10{^ 6}} ^{9.64\hspace{0.07cm} 10{^ 8}}$  &3.82 $10{^8}_{4.45\hspace{0.07cm} 10{^ 6}} ^{5.19\hspace{0.07cm} 10{^ 8}}$  &0.057 \\ \hline
11.7, 2.0, 0.7, 0.3 &1.17 $10{^9}_{7.85\hspace{0.07cm} 10{^ 6}} ^{9.57\hspace{0.07cm} 10{^ 9}}$  &1.84 $10{^9}_{1.24\hspace{0.07cm} 10{^ 9}} ^{2.39\hspace{0.07cm} 10{^ 9}}$  &2.03 $10{^9}_{1.37\hspace{0.07cm} 10{^ 9}} ^{2.64\hspace{0.07cm} 10{^ 9}}$  &1.71 $10{^9}_{1.15\hspace{0.07cm} 10{^ 9}}^{2.22\hspace{0.07cm} 10{^ 9}}$  &1.82 $10{^9}_{9.42\hspace{0.07cm} 10{^ 8}} ^{2.37\hspace{0.07cm} 10{^ 9}}$  &7.46 $10{^8}_{5.73\hspace{0.07cm} 10{^ 6}} ^{1.64\hspace{0.07cm} 10{^ 9}}$  &4.35 $10{^8}_{4.35\hspace{0.07cm} 10{^ 6}} ^{8.39\hspace{0.07cm} 10{^ 8}}$  &0.057 \\ \hline
11.7, 2.0, 0.0, 1.0 &5.24 $10{^8}_{5.24\hspace{0.07cm} 10{^ 6}} ^{6.38\hspace{0.07cm} 10{^ 9}}$  &1.75 $10{^9}_{1.18\hspace{0.07cm} 10{^ 9}} ^{2.28\hspace{0.07cm} 10{^ 9}}$  &1.96 $10{^9}_{1.32\hspace{0.07cm} 10{^ 9}} ^{2.54\hspace{0.07cm} 10{^ 9}}$  &1.65 $10{^9}_{1.11\hspace{0.07cm} 10{^ 9}}^{2.15\hspace{0.07cm} 10{^ 9}}$  &1.79 $10{^9}_{1.06\hspace{0.07cm} 10{^ 9}} ^{2.33\hspace{0.07cm} 10{^ 9}}$  &7.34 $10{^8}_{6.44\hspace{0.07cm} 10{^ 6}} ^{1.42\hspace{0.07cm} 10{^ 9}}$  &4.62 $10{^8}_{5.27\hspace{0.07cm} 10{^ 6}} ^{8.91\hspace{0.07cm} 10{^ 8}}$  &0.056 \\ \hline
10.8, 2.0, 0.0, 0.3 &6.61 $10{^8}_{6.12e+05} ^{5.67\hspace{0.07cm} 10{^ 9}}$  &2.11 $10{^9}_{1.33\hspace{0.07cm} 10{^ 9}} ^{2.66\hspace{0.07cm} 10{^ 9}}$  &1.65 $10{^9}_{9.64\hspace{0.07cm} 10{^ 8}} ^{2.42\hspace{0.07cm} 10{^ 9}}$  &1.14 $10{^9}_{7.78\hspace{0.07cm} 10{^ 8}}^{1.68\hspace{0.07cm} 10{^ 9}}$  &1.06 $10{^9}_{5.74\hspace{0.07cm} 10{^ 8}} ^{1.44\hspace{0.07cm} 10{^ 9}}$  &3.72 $10{^8}_{3.72\hspace{0.07cm} 10{^ 6}} ^{8.01\hspace{0.07cm} 10{^ 8}}$  &2.15 $10{^8}_{2.32\hspace{0.07cm} 10{^ 6}} ^{3.97\hspace{0.07cm} 10{^ 8}}$  &0.065 \\ \hline
10.8, 2.0, 0.7, 0.3 &7.88 $10{^8}_{9.19\hspace{0.07cm} 10{^ 6}} ^{7.88\hspace{0.07cm} 10{^ 9}}$  &2.69 $10{^9}_{1.70\hspace{0.07cm} 10{^ 9}} ^{3.39\hspace{0.07cm} 10{^ 9}}$  &2.26 $10{^9}_{1.42\hspace{0.07cm} 10{^ 9}} ^{3.31\hspace{0.07cm} 10{^ 9}}$  &1.61 $10{^9}_{1.09\hspace{0.07cm} 10{^ 9}}^{2.36\hspace{0.07cm} 10{^ 9}}$  &1.56 $10{^9}_{8.44\hspace{0.07cm} 10{^ 8}} ^{2.12\hspace{0.07cm} 10{^ 9}}$  &5.62 $10{^8}_{5.62\hspace{0.07cm} 10{^ 6}} ^{1.21\hspace{0.07cm} 10{^ 9}}$  &3.65 $10{^8}_{3.95\hspace{0.07cm} 10{^ 6}} ^{6.75\hspace{0.07cm} 10{^ 8}}$  &0.065 \\ \hline
10.8, 2.0, 0.0, 1.0 &1.04 $10{^9}_{7.98\hspace{0.07cm} 10{^ 6}} ^{5.74\hspace{0.07cm} 10{^ 9}}$  &2.48 $10{^9}_{1.67\hspace{0.07cm} 10{^ 9}} ^{3.23\hspace{0.07cm} 10{^ 9}}$  &2.13 $10{^9}_{1.44\hspace{0.07cm} 10{^ 9}} ^{2.77\hspace{0.07cm} 10{^ 9}}$  &1.55 $10{^9}_{1.05\hspace{0.07cm} 10{^ 9}}^{2.31\hspace{0.07cm} 10{^ 9}}$  &1.50 $10{^9}_{7.79\hspace{0.07cm} 10{^ 8}} ^{1.96\hspace{0.07cm} 10{^ 9}}$  &5.81 $10{^8}_{5.81\hspace{0.07cm} 10{^ 6}} ^{1.12\hspace{0.07cm} 10{^ 9}}$  &3.49 $10{^8}_{4.54\hspace{0.07cm} 10{^ 6}} ^{6.74\hspace{0.07cm} 10{^ 8}}$  &0.062 \\ \hline
12.5, 1.7, 0.0, 0.3 &9.48 $10{^8}_{5.98\hspace{0.07cm} 10{^ 6}} ^{7.53\hspace{0.07cm} 10{^ 9}}$  &1.29 $10{^9}_{7.52\hspace{0.07cm} 10{^ 8}} ^{1.62\hspace{0.07cm} 10{^ 9}}$  &1.52 $10{^9}_{1.03\hspace{0.07cm} 10{^ 9}} ^{2.06\hspace{0.07cm} 10{^ 9}}$  &1.27 $10{^9}_{8.65\hspace{0.07cm} 10{^ 8}}^{1.73\hspace{0.07cm} 10{^ 9}}$  &1.30 $10{^9}_{7.01\hspace{0.07cm} 10{^ 8}} ^{1.76\hspace{0.07cm} 10{^ 9}}$  &4.92 $10{^8}_{4.92\hspace{0.07cm} 10{^ 6}} ^{1.06\hspace{0.07cm} 10{^ 9}}$  &3.32 $10{^8}_{3.59\hspace{0.07cm} 10{^ 6}} ^{5.69\hspace{0.07cm} 10{^ 8}}$  &0.054 \\ \hline
12.5, 1.7, 0.7, 0.3 &2.58 $10{^9}_{6.00\hspace{0.07cm} 10{^ 6}} ^{1.20\hspace{0.07cm} 10^{10}}$  &1.54 $10{^9}_{9.02\hspace{0.07cm} 10{^ 8}} ^{2.10\hspace{0.07cm} 10{^ 9}}$  &2.01 $10{^9}_{1.37\hspace{0.07cm} 10{^ 9}} ^{2.73\hspace{0.07cm} 10{^ 9}}$  &1.76 $10{^9}_{1.11\hspace{0.07cm} 10{^ 9}}^{2.58\hspace{0.07cm} 10{^ 9}}$  &1.81 $10{^9}_{9.79\hspace{0.07cm} 10{^ 8}} ^{2.66\hspace{0.07cm} 10{^ 9}}$  &8.55 $10{^8}_{8.55\hspace{0.07cm} 10{^ 6}} ^{1.71\hspace{0.07cm} 10{^ 9}}$  &4.78 $10{^8}_{4.42\hspace{0.07cm} 10{^ 6}} ^{9.53\hspace{0.07cm} 10{^ 8}}$  &0.051 \\ \hline
12.5, 1.7, 0.0, 1.0 &4.31 $10{^8}_{2.00\hspace{0.07cm} 10{^7}} ^{7.37\hspace{0.07cm} 10{^ 9}}$  &1.54 $10{^9}_{9.72\hspace{0.07cm} 10{^ 8}} ^{2.09\hspace{0.07cm} 10{^ 9}}$  &2.02 $10{^9}_{1.28\hspace{0.07cm} 10{^ 9}} ^{2.75\hspace{0.07cm} 10{^ 9}}$  &1.72 $10{^9}_{1.17\hspace{0.07cm} 10{^ 9}}^{2.52\hspace{0.07cm} 10{^ 9}}$  &1.86 $10{^9}_{1.01\hspace{0.07cm} 10{^ 9}} ^{2.53\hspace{0.07cm} 10{^ 9}}$  &8.18 $10{^8}_{8.18\hspace{0.07cm} 10{^ 6}} ^{1.63\hspace{0.07cm} 10{^ 9}}$  &5.25 $10{^8}_{5.25\hspace{0.07cm} 10{^ 6}} ^{9.71\hspace{0.07cm} 10{^ 8}}$  &0.054 \\ \hline
11.7, 1.7, 0.0, 0.3 &1.03 $10{^9}_{7.59\hspace{0.07cm} 10{^ 6}} ^{6.51\hspace{0.07cm} 10{^ 9}}$  &1.60 $10{^9}_{1.01\hspace{0.07cm} 10{^ 9}} ^{2.01\hspace{0.07cm} 10{^ 9}}$  &1.59 $10{^9}_{1.00\hspace{0.07cm} 10{^ 9}} ^{2.16\hspace{0.07cm} 10{^ 9}}$  &1.17 $10{^9}_{7.98\hspace{0.07cm} 10{^ 8}}^{1.72\hspace{0.07cm} 10{^ 9}}$  &1.11 $10{^9}_{5.99\hspace{0.07cm} 10{^ 8}} ^{1.51\hspace{0.07cm} 10{^ 9}}$  &3.78 $10{^8}_{3.50\hspace{0.07cm} 10{^ 6}} ^{8.14\hspace{0.07cm} 10{^ 8}}$  &2.69 $10{^8}_{3.65\hspace{0.07cm} 10{^ 6}} ^{4.26\hspace{0.07cm} 10{^ 8}}$  &0.060 \\ \hline
11.7, 1.7, 0.7, 0.3 &8.92 $10{^8}_{6.01\hspace{0.07cm} 10{^ 6}} ^{8.35\hspace{0.07cm} 10{^ 9}}$  &2.11 $10{^9}_{1.43\hspace{0.07cm} 10{^ 9}} ^{2.75\hspace{0.07cm} 10{^ 9}}$  &2.12 $10{^9}_{1.43\hspace{0.07cm} 10{^ 9}} ^{2.75\hspace{0.07cm} 10{^ 9}}$  &1.64 $10{^9}_{1.11\hspace{0.07cm} 10{^ 9}}^{2.14\hspace{0.07cm} 10{^ 9}}$  &1.61 $10{^9}_{8.36\hspace{0.07cm} 10{^ 8}} ^{2.10\hspace{0.07cm} 10{^ 9}}$  &6.37 $10{^8}_{6.37\hspace{0.07cm} 10{^ 6}} ^{1.23\hspace{0.07cm} 10{^ 9}}$  &3.82 $10{^8}_{4.36\hspace{0.07cm} 10{^ 6}} ^{7.38\hspace{0.07cm} 10{^ 8}}$  &0.059 \\ \hline
11.7, 1.7, 0.0, 1.0 &1.22 $10{^9}_{9.41\hspace{0.07cm} 10{^ 6}} ^{6.77\hspace{0.07cm} 10{^ 9}}$  &1.90 $10{^9}_{1.12\hspace{0.07cm} 10{^ 9}} ^{2.47\hspace{0.07cm} 10{^ 9}}$  &2.03 $10{^9}_{1.37\hspace{0.07cm} 10{^ 9}} ^{2.64\hspace{0.07cm} 10{^ 9}}$  &1.60 $10{^9}_{1.08\hspace{0.07cm} 10{^ 9}}^{2.08\hspace{0.07cm} 10{^ 9}}$  &1.61 $10{^9}_{8.33\hspace{0.07cm} 10{^ 8}} ^{2.09\hspace{0.07cm} 10{^ 9}}$  &6.07 $10{^8}_{6.07\hspace{0.07cm} 10{^ 6}} ^{1.34\hspace{0.07cm} 10{^ 9}}$  &4.21 $10{^8}_{4.80\hspace{0.07cm} 10{^ 6}} ^{7.13\hspace{0.07cm} 10{^ 8}}$  &0.057 \\ \hline
10.8, 1.7, 0.0, 0.3 &8.92 $10{^8}_{8.27\hspace{0.07cm} 10{^ 6}} ^{5.63\hspace{0.07cm} 10{^ 9}}$  &2.43 $10{^9}_{1.53\hspace{0.07cm} 10{^ 9}} ^{3.05\hspace{0.07cm} 10{^ 9}}$  &1.71 $10{^9}_{9.24\hspace{0.07cm} 10{^ 8}} ^{2.70\hspace{0.07cm} 10{^ 9}}$  &1.13 $10{^9}_{7.71\hspace{0.07cm} 10{^ 8}}^{1.66\hspace{0.07cm} 10{^ 9}}$  &9.98 $10{^8}_{5.40\hspace{0.07cm} 10{^ 8}} ^{1.36\hspace{0.07cm} 10{^ 9}}$  &3.17 $10{^8}_{2.00\hspace{0.07cm} 10{^ 6}} ^{6.83\hspace{0.07cm} 10{^ 8}}$  &1.72 $10{^8}_{2.00\hspace{0.07cm} 10{^ 6}} ^{3.17\hspace{0.07cm} 10{^ 8}}$  &0.069 \\ \hline
10.8, 1.7, 0.7, 0.3 &2.28 $10{^9}_{1.55\hspace{0.07cm} 10{^7}} ^{8.40\hspace{0.07cm} 10{^ 9}}$  &2.77 $10{^9}_{1.75\hspace{0.07cm} 10{^ 9}} ^{3.77\hspace{0.07cm} 10{^ 9}}$  &2.45 $10{^9}_{1.43\hspace{0.07cm} 10{^ 9}} ^{3.60\hspace{0.07cm} 10{^ 9}}$  &1.50 $10{^9}_{1.02\hspace{0.07cm} 10{^ 9}}^{2.38\hspace{0.07cm} 10{^ 9}}$  &1.40 $10{^9}_{7.57\hspace{0.07cm} 10{^ 8}} ^{1.90\hspace{0.07cm} 10{^ 9}}$  &5.11 $10{^8}_{5.11\hspace{0.07cm} 10{^ 6}} ^{1.10\hspace{0.07cm} 10{^ 9}}$  &2.90 $10{^8}_{2.90\hspace{0.07cm} 10{^ 6}} ^{5.35\hspace{0.07cm} 10{^ 8}}$  &0.065 \\ \hline
10.8, 1.7, 0.0, 1.0 &1.06 $10{^9}_{9.80\hspace{0.07cm} 10{^ 6}} ^{5.73\hspace{0.07cm} 10{^ 9}}$  &2.86 $10{^9}_{1.81\hspace{0.07cm} 10{^ 9}} ^{3.60\hspace{0.07cm} 10{^ 9}}$  &2.27 $10{^9}_{1.32\hspace{0.07cm} 10{^ 9}} ^{3.33\hspace{0.07cm} 10{^ 9}}$  &1.54 $10{^9}_{1.05\hspace{0.07cm} 10{^ 9}}^{2.26\hspace{0.07cm} 10{^ 9}}$  &1.41 $10{^9}_{7.61\hspace{0.07cm} 10{^ 8}} ^{1.91\hspace{0.07cm} 10{^ 9}}$  &4.90 $10{^8}_{4.54\hspace{0.07cm} 10{^ 6}} ^{1.06\hspace{0.07cm} 10{^ 9}}$  &3.07 $10{^8}_{3.87\hspace{0.07cm} 10{^ 6}} ^{5.26\hspace{0.07cm} 10{^ 8}}$  &0.066 \\ \hline
12.5, 1.5, 0.0, 0.3 &1.31 $10{^9}_{7.09\hspace{0.07cm} 10{^ 6}} ^{7.65\hspace{0.07cm} 10{^ 9}}$  &1.25 $10{^9}_{7.88\hspace{0.07cm} 10{^ 8}} ^{1.70\hspace{0.07cm} 10{^ 9}}$  &1.62 $10{^9}_{1.02\hspace{0.07cm} 10{^ 9}} ^{2.20\hspace{0.07cm} 10{^ 9}}$  &1.24 $10{^9}_{8.42\hspace{0.07cm} 10{^ 8}}^{1.81\hspace{0.07cm} 10{^ 9}}$  &1.19 $10{^9}_{6.44\hspace{0.07cm} 10{^ 8}} ^{1.75\hspace{0.07cm} 10{^ 9}}$  &4.28 $10{^8}_{3.96\hspace{0.07cm} 10{^ 6}} ^{9.22\hspace{0.07cm} 10{^ 8}}$  &3.13 $10{^8}_{3.65\hspace{0.07cm} 10{^ 6}} ^{4.60\hspace{0.07cm} 10{^ 8}}$  &0.058 \\ \hline
12.5, 1.5, 0.7, 0.3 &8.95 $10{^8}_{7.84\hspace{0.07cm} 10{^ 6}} ^{8.38\hspace{0.07cm} 10{^ 9}}$  &1.82 $10{^9}_{1.23\hspace{0.07cm} 10{^ 9}} ^{2.37\hspace{0.07cm} 10{^ 9}}$  &2.09 $10{^9}_{1.41\hspace{0.07cm} 10{^ 9}} ^{2.71\hspace{0.07cm} 10{^ 9}}$  &1.70 $10{^9}_{1.14\hspace{0.07cm} 10{^ 9}}^{2.21\hspace{0.07cm} 10{^ 9}}$  &1.86 $10{^9}_{9.64\hspace{0.07cm} 10{^ 8}} ^{2.42\hspace{0.07cm} 10{^ 9}}$  &6.58 $10{^8}_{5.77\hspace{0.07cm} 10{^ 6}} ^{1.45\hspace{0.07cm} 10{^ 9}}$  &4.90 $10{^8}_{4.90\hspace{0.07cm} 10{^ 6}} ^{8.29\hspace{0.07cm} 10{^ 8}}$  &0.057 \\ \hline
12.5, 1.5, 0.0, 1.0 &5.64 $10{^8}_{8.36\hspace{0.07cm} 10{^ 6}} ^{6.02\hspace{0.07cm} 10{^ 9}}$  &1.73 $10{^9}_{1.02\hspace{0.07cm} 10{^ 9}} ^{2.25\hspace{0.07cm} 10{^ 9}}$  &1.99 $10{^9}_{1.34\hspace{0.07cm} 10{^ 9}} ^{2.59\hspace{0.07cm} 10{^ 9}}$  &1.66 $10{^9}_{1.12\hspace{0.07cm} 10{^ 9}}^{2.46\hspace{0.07cm} 10{^ 9}}$  &1.77 $10{^9}_{1.04\hspace{0.07cm} 10{^ 9}} ^{2.30\hspace{0.07cm} 10{^ 9}}$  &7.22 $10{^8}_{6.33\hspace{0.07cm} 10{^ 6}} ^{1.39\hspace{0.07cm} 10{^ 9}}$  &4.60 $10{^8}_{5.25\hspace{0.07cm} 10{^ 6}} ^{7.79\hspace{0.07cm} 10{^ 8}}$  &0.056 \\ \hline
11.7, 1.5, 0.0, 0.3 &1.26 $10{^9}_{7.97\hspace{0.07cm} 10{^ 6}} ^{5.86\hspace{0.07cm} 10{^ 9}}$  &1.68 $10{^9}_{1.06\hspace{0.07cm} 10{^ 9}} ^{2.12\hspace{0.07cm} 10{^ 9}}$  &1.72 $10{^9}_{1.00\hspace{0.07cm} 10{^ 9}} ^{2.52\hspace{0.07cm} 10{^ 9}}$  &1.10 $10{^9}_{8.06\hspace{0.07cm} 10{^ 8}}^{1.74\hspace{0.07cm} 10{^ 9}}$  &1.09 $10{^9}_{5.49\hspace{0.07cm} 10{^ 8}} ^{1.49\hspace{0.07cm} 10{^ 9}}$  &3.22 $10{^8}_{2.76\hspace{0.07cm} 10{^ 6}} ^{7.48\hspace{0.07cm} 10{^ 8}}$  &2.46 $10{^8}_{2.65\hspace{0.07cm} 10{^ 6}} ^{3.61\hspace{0.07cm} 10{^ 8}}$  &0.065 \\ \hline
11.7, 1.5, 0.7, 0.3 &1.35 $10{^9}_{1.25\hspace{0.07cm} 10{^7}} ^{7.90\hspace{0.07cm} 10{^ 9}}$  &2.28 $10{^9}_{1.44\hspace{0.07cm} 10{^ 9}} ^{2.87\hspace{0.07cm} 10{^ 9}}$  &2.19 $10{^9}_{1.38\hspace{0.07cm} 10{^ 9}} ^{2.98\hspace{0.07cm} 10{^ 9}}$  &1.59 $10{^9}_{1.08\hspace{0.07cm} 10{^ 9}}^{2.33\hspace{0.07cm} 10{^ 9}}$  &1.55 $10{^9}_{8.41\hspace{0.07cm} 10{^ 8}} ^{2.11\hspace{0.07cm} 10{^ 9}}$  &5.50 $10{^8}_{5.50\hspace{0.07cm} 10{^ 6}} ^{1.19\hspace{0.07cm} 10{^ 9}}$  &3.52 $10{^8}_{3.52\hspace{0.07cm} 10{^ 6}} ^{6.02\hspace{0.07cm} 10{^ 8}}$  &0.061 \\ \hline
11.7, 1.5, 0.0, 1.0 &1.10 $10{^9}_{7.38\hspace{0.07cm} 10{^ 6}} ^{6.91\hspace{0.07cm} 10{^ 9}}$  &2.05 $10{^9}_{1.38\hspace{0.07cm} 10{^ 9}} ^{2.66\hspace{0.07cm} 10{^ 9}}$  &2.11 $10{^9}_{1.42\hspace{0.07cm} 10{^ 9}} ^{2.74\hspace{0.07cm} 10{^ 9}}$  &1.57 $10{^9}_{1.06\hspace{0.07cm} 10{^ 9}}^{2.34\hspace{0.07cm} 10{^ 9}}$  &1.53 $10{^9}_{7.95\hspace{0.07cm} 10{^ 8}} ^{2.00\hspace{0.07cm} 10{^ 9}}$  &5.01 $10{^8}_{5.71\hspace{0.07cm} 10{^ 6}} ^{1.10\hspace{0.07cm} 10{^ 9}}$  &4.24 $10{^8}_{4.24\hspace{0.07cm} 10{^ 6}} ^{6.29\hspace{0.07cm} 10{^ 8}}$  &0.060 \\ \hline
10.8, 1.5, 0.0, 0.3 &1.40 $10{^9}_{7.56\hspace{0.07cm} 10{^ 6}} ^{6.01\hspace{0.07cm} 10{^ 9}}$  &2.60 $10{^9}_{1.52\hspace{0.07cm} 10{^ 9}} ^{3.27\hspace{0.07cm} 10{^ 9}}$  &1.79 $10{^9}_{9.69\hspace{0.07cm} 10{^ 8}} ^{2.84\hspace{0.07cm} 10{^ 9}}$  &1.07 $10{^9}_{7.32\hspace{0.07cm} 10{^ 8}}^{1.70\hspace{0.07cm} 10{^ 9}}$  &9.61 $10{^8}_{4.82\hspace{0.07cm} 10{^ 8}} ^{1.31\hspace{0.07cm} 10{^ 9}}$  &2.62 $10{^8}_{2.25\hspace{0.07cm} 10{^ 6}} ^{6.09\hspace{0.07cm} 10{^ 8}}$  &1.81 $10{^8}_{2.28\hspace{0.07cm} 10{^ 6}} ^{2.87\hspace{0.07cm} 10{^ 8}}$  &0.070 \\ \hline
10.8, 1.5, 0.7, 0.3 &1.08 $10{^9}_{9.96\hspace{0.07cm} 10{^ 6}} ^{7.33\hspace{0.07cm} 10{^ 9}}$  &3.40 $10{^9}_{2.14\hspace{0.07cm} 10{^ 9}} ^{4.28\hspace{0.07cm} 10{^ 9}}$  &2.53 $10{^9}_{1.48\hspace{0.07cm} 10{^ 9}} ^{3.72\hspace{0.07cm} 10{^ 9}}$  &1.59 $10{^9}_{1.08\hspace{0.07cm} 10{^ 9}}^{2.34\hspace{0.07cm} 10{^ 9}}$  &1.37 $10{^9}_{7.43\hspace{0.07cm} 10{^ 8}} ^{1.87\hspace{0.07cm} 10{^ 9}}$  &4.54 $10{^8}_{4.21\hspace{0.07cm} 10{^ 6}} ^{9.78\hspace{0.07cm} 10{^ 8}}$  &2.77 $10{^8}_{3.23\hspace{0.07cm} 10{^ 6}} ^{4.74\hspace{0.07cm} 10{^ 8}}$  &0.071 \\ \hline
10.8 , 1.5, 0.0, 1.0 &1.03 $10{^9}_{9.54\hspace{0.07cm} 10{^ 6}} ^{5.58\hspace{0.07cm} 10{^ 9}}$  &3.16 $10{^9}_{1.99\hspace{0.07cm} 10{^ 9}} ^{3.98\hspace{0.07cm} 10{^ 9}}$  &2.39 $10{^9}_{1.40\hspace{0.07cm} 10{^ 9}} ^{3.51\hspace{0.07cm} 10{^ 9}}$  &1.53 $10{^9}_{1.04\hspace{0.07cm} 10{^ 9}}^{2.25\hspace{0.07cm} 10{^ 9}}$  &1.38 $10{^9}_{6.90\hspace{0.07cm} 10{^ 8}} ^{1.87\hspace{0.07cm} 10{^ 9}}$  &4.26 $10{^8}_{3.95\hspace{0.07cm} 10{^ 6}} ^{9.18\hspace{0.07cm} 10{^ 8}}$  &2.91 $10{^8}_{3.15\hspace{0.07cm} 10{^ 6}} ^{4.62\hspace{0.07cm} 10{^ 8}}$  &0.069 \\ \hline
12.5, 1.3, 0.0, 0.3 &1.69 $10{^9}_{1.24\hspace{0.07cm} 10{^7}} ^{7.85\hspace{0.07cm} 10{^ 9}}$  &1.39 $10{^9}_{8.14\hspace{0.07cm} 10{^ 8}} ^{1.89\hspace{0.07cm} 10{^ 9}}$  &1.57 $10{^9}_{9.91\hspace{0.07cm} 10{^ 8}} ^{2.31\hspace{0.07cm} 10{^ 9}}$  &1.20 $10{^9}_{7.56\hspace{0.07cm} 10{^ 8}}^{1.76\hspace{0.07cm} 10{^ 9}}$  &1.16 $10{^9}_{5.81\hspace{0.07cm} 10{^ 8}} ^{1.58\hspace{0.07cm} 10{^ 9}}$  &3.86 $10{^8}_{3.86\hspace{0.07cm} 10{^ 6}} ^{8.32\hspace{0.07cm} 10{^ 8}}$  &2.07 $10{^8}_{2.41\hspace{0.07cm} 10{^ 6}} ^{4.12\hspace{0.07cm} 10{^ 8}}$  &0.057 \\ \hline
12.5, 1.3, 0.7, 0.3 &1.16 $10{^9}_{6.84\hspace{0.07cm} 10{^ 6}} ^{8.34\hspace{0.07cm} 10{^ 9}}$  &1.93 $10{^9}_{1.30\hspace{0.07cm} 10{^ 9}} ^{2.51\hspace{0.07cm} 10{^ 9}}$  &2.14 $10{^9}_{1.44\hspace{0.07cm} 10{^ 9}} ^{2.78\hspace{0.07cm} 10{^ 9}}$  &1.69 $10{^9}_{1.14\hspace{0.07cm} 10{^ 9}}^{2.20\hspace{0.07cm} 10{^ 9}}$  &1.67 $10{^9}_{8.63\hspace{0.07cm} 10{^ 8}} ^{2.17\hspace{0.07cm} 10{^ 9}}$  &6.12 $10{^8}_{6.12\hspace{0.07cm} 10{^ 6}} ^{1.35\hspace{0.07cm} 10{^ 9}}$  &4.03 $10{^8}_{4.59\hspace{0.07cm} 10{^ 6}} ^{6.82\hspace{0.07cm} 10{^ 8}}$  &0.057 \\ \hline
12.5, 1.3, 0.0, 1.0 &1.23 $10{^9}_{8.70\hspace{0.07cm} 10{^ 6}} ^{7.32\hspace{0.07cm} 10{^ 9}}$  &1.76 $10{^9}_{1.05\hspace{0.07cm} 10{^ 9}} ^{2.35\hspace{0.07cm} 10{^ 9}}$  &2.07 $10{^9}_{1.39\hspace{0.07cm} 10{^ 9}} ^{2.77\hspace{0.07cm} 10{^ 9}}$  &1.61 $10{^9}_{1.08\hspace{0.07cm} 10{^ 9}}^{2.28\hspace{0.07cm} 10{^ 9}}$  &1.66 $10{^9}_{8.82\hspace{0.07cm} 10{^ 8}} ^{2.35\hspace{0.07cm} 10{^ 9}}$  &5.52 $10{^8}_{5.21\hspace{0.07cm} 10{^ 6}} ^{1.24\hspace{0.07cm} 10{^ 9}}$  &4.70 $10{^8}_{5.58\hspace{0.07cm} 10{^ 6}} ^{6.63\hspace{0.07cm} 10{^ 8}}$  &0.056 \\ \hline
11.7, 1.3, 0.0, 0.3 &1.57 $10{^9}_{6.26\hspace{0.07cm} 10{^ 6}} ^{7.30\hspace{0.07cm} 10{^ 9}}$  &1.76 $10{^9}_{1.11\hspace{0.07cm} 10{^ 9}} ^{2.39\hspace{0.07cm} 10{^ 9}}$  &1.73 $10{^9}_{1.01\hspace{0.07cm} 10{^ 9}} ^{2.54\hspace{0.07cm} 10{^ 9}}$  &1.13 $10{^9}_{7.67\hspace{0.07cm} 10{^ 8}}^{1.65\hspace{0.07cm} 10{^ 9}}$  &9.44 $10{^8}_{5.11\hspace{0.07cm} 10{^ 8}} ^{1.39\hspace{0.07cm} 10{^ 9}}$  &3.29 $10{^8}_{4.15\hspace{0.07cm} 10{^ 6}} ^{7.10\hspace{0.07cm} 10{^ 8}}$  &1.79 $10{^8}_{1.93\hspace{0.07cm} 10{^ 6}} ^{3.06\hspace{0.07cm} 10{^ 8}}$  &0.063 \\ \hline
11.7, 1.3, 0.7, 0.3 &1.42 $10{^9}_{1.32\hspace{0.07cm} 10{^7}} ^{8.30\hspace{0.07cm} 10{^ 9}}$  &2.46 $10{^9}_{1.55\hspace{0.07cm} 10{^ 9}} ^{3.10\hspace{0.07cm} 10{^ 9}}$  &2.32 $10{^9}_{1.46\hspace{0.07cm} 10{^ 9}} ^{3.15\hspace{0.07cm} 10{^ 9}}$  &1.57 $10{^9}_{1.07\hspace{0.07cm} 10{^ 9}}^{2.31\hspace{0.07cm} 10{^ 9}}$  &1.49 $10{^9}_{8.05\hspace{0.07cm} 10{^ 8}} ^{2.02\hspace{0.07cm} 10{^ 9}}$  &4.87 $10{^8}_{4.87\hspace{0.07cm} 10{^ 6}} ^{1.05\hspace{0.07cm} 10{^ 9}}$  &3.19 $10{^8}_{2.96\hspace{0.07cm} 10{^ 6}} ^{5.46\hspace{0.07cm} 10{^ 8}}$  &0.063 \\ \hline
11.7, 1.3, 0.0, 1.0 &1.28 $10{^9}_{9.09\hspace{0.07cm} 10{^ 6}} ^{6.82\hspace{0.07cm} 10{^ 9}}$  &2.23 $10{^9}_{1.33\hspace{0.07cm} 10{^ 9}} ^{2.81\hspace{0.07cm} 10{^ 9}}$  &2.21 $10{^9}_{1.40\hspace{0.07cm} 10{^ 9}} ^{3.13\hspace{0.07cm} 10{^ 9}}$  &1.54 $10{^9}_{1.03\hspace{0.07cm} 10{^ 9}}^{2.30\hspace{0.07cm} 10{^ 9}}$  &1.46 $10{^9}_{7.30\hspace{0.07cm} 10{^ 8}} ^{2.06\hspace{0.07cm} 10{^ 9}}$  &4.74 $10{^8}_{4.74\hspace{0.07cm} 10{^ 6}} ^{1.06\hspace{0.07cm} 10{^ 9}}$  &3.11 $10{^8}_{3.69\hspace{0.07cm} 10{^ 6}} ^{5.22\hspace{0.07cm} 10{^ 8}}$  &0.061 \\ \hline
10.8, 1.3, 0.0, 0.3 &1.47 $10{^9}_{1.10\hspace{0.07cm} 10{^7}} ^{5.52\hspace{0.07cm} 10{^ 9}}$  &2.81 $10{^9}_{1.67\hspace{0.07cm} 10{^ 9}} ^{3.54\hspace{0.07cm} 10{^ 9}}$  &1.95 $10{^9}_{1.03\hspace{0.07cm} 10{^ 9}} ^{3.08\hspace{0.07cm} 10{^ 9}}$  &1.11 $10{^9}_{7.44\hspace{0.07cm} 10{^ 8}}^{1.66\hspace{0.07cm} 10{^ 9}}$  &8.85 $10{^8}_{4.44\hspace{0.07cm} 10{^ 8}} ^{1.25\hspace{0.07cm} 10{^ 9}}$  &2.28 $10{^8}_{2.56\hspace{0.07cm} 10{^ 6}} ^{5.42\hspace{0.07cm} 10{^ 8}}$  &1.64 $10{^8}_{2.19\hspace{0.07cm} 10{^ 6}} ^{2.45\hspace{0.07cm} 10{^ 8}}$  &0.071 \\ \hline
10.8, 1.3, 0.7, 0.3 & 1.63 $10{^9}_{1.30\hspace{0.07cm} 10{^7}} ^{8.17\hspace{0.07cm} 10{^ 9}}$  &3.70 $10{^9}_{2.33\hspace{0.07cm} 10{^ 9}} ^{4.65\hspace{0.07cm} 10{^ 9}}$  &2.63 $10{^9}_{1.42\hspace{0.07cm} 10{^ 9}} ^{3.86\hspace{0.07cm} 10{^ 9}}$  &1.58 $10{^9}_{1.08\hspace{0.07cm} 10{^ 9}}^{2.32\hspace{0.07cm} 10{^ 9}}$  &1.35 $10{^9}_{6.77\hspace{0.07cm} 10{^ 8}} ^{1.84\hspace{0.07cm} 10{^ 9}}$  &3.69 $10{^8}_{3.42\hspace{0.07cm} 10{^ 6}} ^{8.58\hspace{0.07cm} 10{^ 8}}$  &2.66 $10{^8}_{3.35\hspace{0.07cm} 10{^ 6}} ^{4.21\hspace{0.07cm} 10{^ 8}}$  &0.072 \\ \hline
10.8, 1.3, 0.0, 1.0 & 8.68 $10{^8}_{6.51\hspace{0.07cm} 10{^ 6}} ^{3.66\hspace{0.07cm} 10{^ 9}}$  &3.50 $10{^9}_{2.21\hspace{0.07cm} 10{^ 9}} ^{4.41\hspace{0.07cm} 10{^ 9}}$  &2.53 $10{^9}_{1.34\hspace{0.07cm} 10{^ 9}} ^{4.00\hspace{0.07cm} 10{^ 9}}$  &1.56 $10{^9}_{1.04\hspace{0.07cm} 10{^ 9}}^{2.33\hspace{0.07cm} 10{^ 9}}$  &1.30 $10{^9}_{6.50\hspace{0.07cm} 10{^ 8}} ^{1.83\hspace{0.07cm} 10{^ 9}}$  &3.75 $10{^8}_{4.45\hspace{0.07cm} 10{^ 6}} ^{8.39\hspace{0.07cm} 10{^ 8}}$  &2.73 $10{^8}_{3.85\hspace{0.07cm} 10{^ 6}}^{4.08\hspace{0.07cm} 10{^ 8}}$  &0.072 \\ \hline
\end{tabular}\\
\end{center}
\end{table*}

\section{\label{source_spectra} Sources of the Cosmic Background and their spectra}

To determine $\varphi(z)$, we need to establish the range of acceptable average SEDs (L$_{\nu^{'}}$) of the sources
that are making the background.\\
Locally Soifer \& Neugebauer (1991) have studied the population of IR galaxies at low 
redshift using the 
IRAS all sky survey. They have established the integrated IR  luminosity per unit volume in the 
Universe today, its SED, the luminosity function of IR galaxies and 
the SED as a function of the integrated IR luminosity. 
The IR luminosity function shows two main components. One is associated 
with normal spiral galaxies which radiate a fraction of their energy 
in the IR (our Galaxy for example radiates about one third of its luminosity 
in the IR). These galaxies 
are expected to have a luminosity function similar to the luminosity function of galaxies in the optical. 
However, the IR luminosity function does not show an exponential cut off like optical galaxies but 
display a power law behaviour 
at high luminosities: $\frac{dN}{dL} = L^{-2}$. This is also shown by more recent compilations of the
IR luminosity functions (Sanders \& Mirabel, 1996). 
This shows that the very luminous IR galaxies cannot be the IR counterpart of optical galaxies. 
In fact they have been shown to be starburst galaxies often 
associated with merging or interacting systems and for which the ratio of IR to optical 
luminosity increases with the bolometric luminosity (Sanders \& Mirabel, 1996; Fang et al. 1998).
One can thus conjecture that the luminosity function of mergers and interacting galaxies is very different 
from the luminosity function of normal galaxies.
Using this rough separation, locally the relative contributions of starburst to normal 
galaxies is less than 10$\%$. The integrated luminosity of normal galaxies is 
dominated by L$_{\star}$ optical galaxies (L$_{IR \star}$ $\simeq$ 10$^{10}$  L$_{\odot}$) whereas the 
integrated luminosity of IR starburst is dominated by galaxies with L$_{IR}\sim$10$^{11}$  
L$_{\odot}$ (see for example Sanders \& Mirabel, 1996). 
\\
26 of the galaxy detected by ISOCAM in the HDF north 
are identified with galaxies with known redshifts. 
Aussel (1999) has shown that the IR galaxies have luminosities greater than
3 10$^{10}$~L${\odot}$ at 8.5 $\mu$m (in the rest frame) with a median redshift of 0.75,
and thus bolometric luminosities between 1 and 3 10$^{11}$~L${\odot}$.
These galaxies, identified mostly as interacting systems or spiral galaxies,
are luminous IR galaxies undergoing a starburst
phase. \\
Spectra of IR galaxies have been modeled by Maffei (1994),
using the observational correlation of the IRAS flux ratio
12/60, 25/60 and 60/100 with the IR luminosity (Soifer \& Neugeubauer, 1991).
Examples of spectra for three different luminosities are shown in Fig. 2.
The luminous IR galaxies are emitting more than 95$\%$ of their energy
in the far-IR. Taking only such kind of galaxies obviously fails
to reproduce the optical Extragalactic Background (EB). That is why we concentrate hereafter
only on the far-IR part of the EB (the so-called Cosmic Far-IR Background,
CFIRB).

It is possible however that part of the CFIRB energy might be due to dust enshrouded Active
Galactic Nuclei (AGNs) for which the far-IR SED is very similar to starburst galaxies of similar luminosity.
Based on the assumptions that 10$\%$ of the mass
accreting into balck hole is turned into energy and that the black hole
masses measured in the HDF (Ford et al. 1998) are typical of galaxies,
the AGN background energy would be in order of 10$\%$ of that from stars
(Eales et al. 1999). These calculations are highly uncertain but are supported
by the recent work of Almaini et al. (1999). 
The SEDs of high-redshift dust enshrouded AGNs are similar to those observed locally, 
and one can explain 10-20 $\%$ of the CFIRB.
Because of the similar far-IR SEDs, the question of the fraction of the CFIRB due to AGNs is not 
relevant to the determination of the IR radiation production history.
The estimates mentioned above are only important because they indicate that the IR radiation 
production history is likely to reflect the star formation history.
Locally the average far-IR SED adjusted on the IRAS data (Soifer \& Neugeubauer, 1991) 
is well represented by the SED of a 8 10$^{10}$~L${\odot}$ IR galaxy. 
As we have seen, at z $\sim $ 0.7 the mid IR production 
is dominated by galaxies typically 2.5 times more luminous. Finally at higher redshift the main 
indication comes from the SCUBA deep surveys and also indicates an average SED dominated by ultraluminous 
IR galaxies. To take into account the uncertainty on the average far-IR SED, we use for 
the determination of the IR radiation production history, the SEDs from Maffei (1994) for galaxies with 
luminosities 7 10$^{10}$~L${\odot}$, 5 10$^{11}$~L${\odot}$ and 3 10$^{12}$~L${\odot}$ which takes 
into account the change in the peak emission wavelength with luminosity. For the long wavelength behaviour, 
we allow the dust emissivity index of the model to vary from 1.3 to 2 which covers generously the uncertainty 
on this parameter for the average SED.

\begin{center}
\begin{figure}  
\epsfxsize=9.cm
\epsfysize=7.cm
\epsfbox{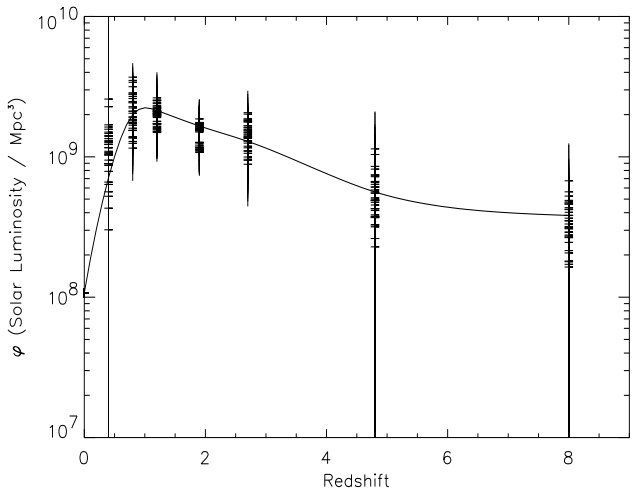}
\caption{ \label{phi_all} Co-moving luminosity density distribution as derived from the CFIRB
for all cases listed in Table 1. Also shown is the best fit passing through all cases.}
\end{figure}
\end{center}

\begin{figure*}
\begin{minipage}{8.3cm}
\epsfxsize=9.cm
\epsfysize=7.cm
\hspace{0.5cm}
\vspace{0.4cm}
\epsfbox{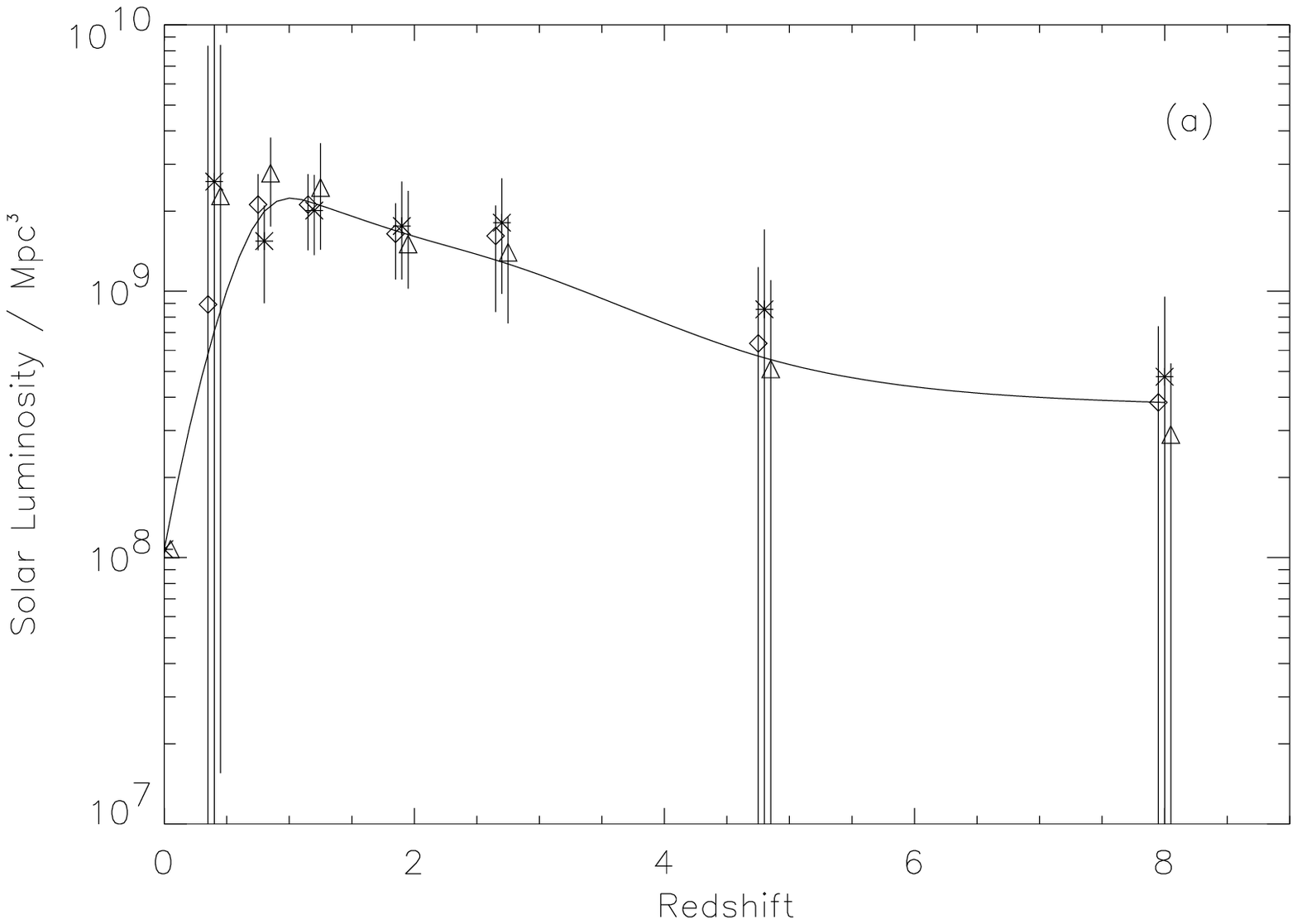}
\end{minipage}
\begin{minipage}{8.3cm}
\epsfxsize=9.cm
\epsfysize=7.cm
\hspace{1.5cm}
\vspace{0.4cm}
\epsfbox{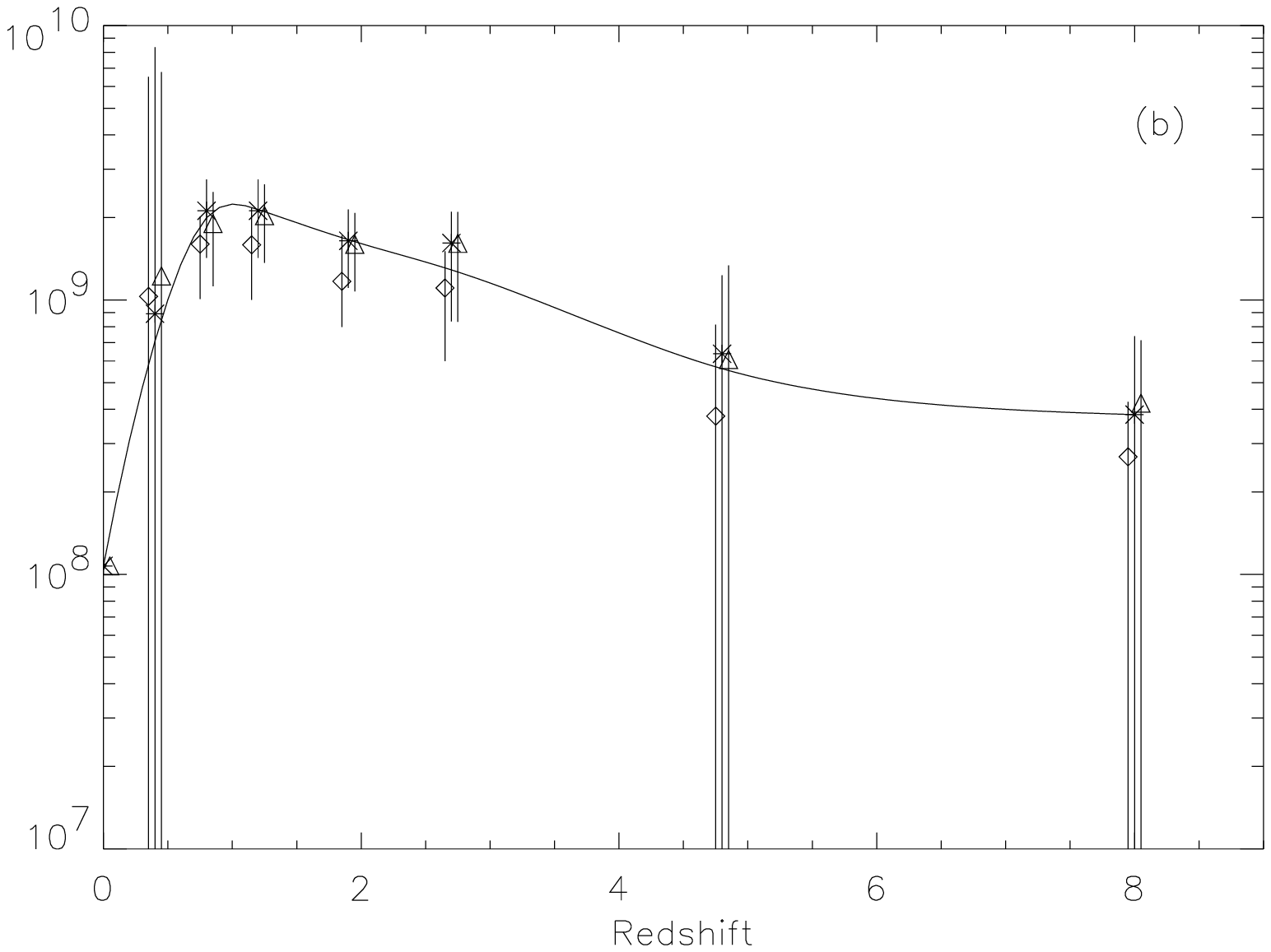}
\end{minipage}\\
\vspace{-1.cm}
\caption{\label{fig_sfr} Co-moving luminosity density distribution as derived from the CFIRB
for different sets of parameters (with a fixed dust spectral index of 1.7).
a: for a given cosmological model ($\Omega_0$=0.3, $\Omega_{\Lambda}$=0.7) and different
luminosities (star: 3~10$^{12}$ L$_{\odot}$, diamond: 5~10$^{11}$ L$_{\odot}$, 
triangle: 7~10$^{10}$ L$_{\odot}$). b: given a luminosity of 5 10$^{11}$ L$_{\odot}$ and three
cosmological models with h=0.65 (star: $\Omega_0$=0.3, $\Omega_{\Lambda}$=0.7, 
diamond: $\Omega_0$=0.3, $\Omega_{\Lambda}$=0, triangle: $\Omega_0$=1, $\Omega_{\Lambda}$=0).
In each plot, points around a redshift value have been arbitrarly shifted
to see the uncertainties.}
\end{figure*}

\begin{figure*}  
\epsfbox{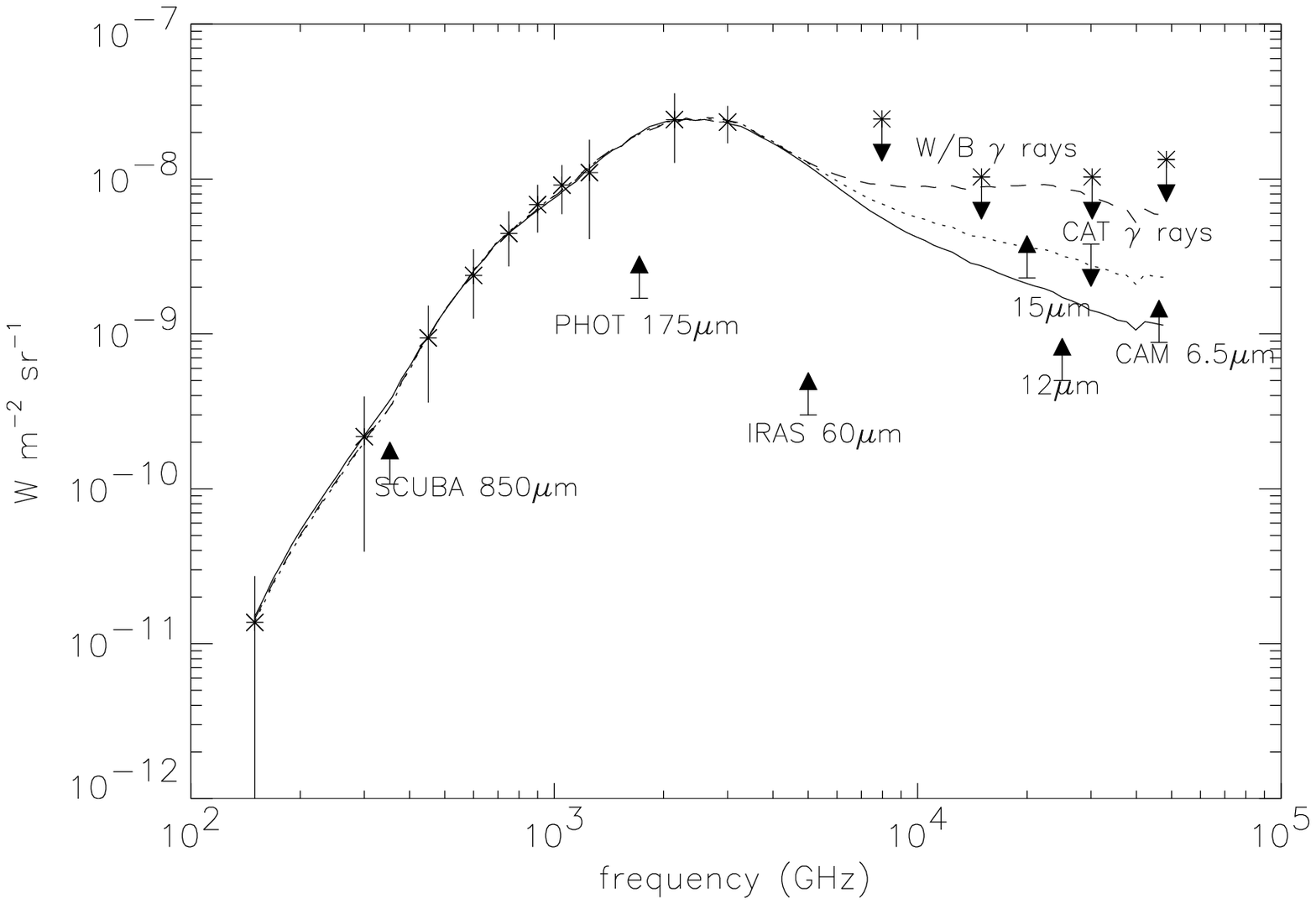}
\caption{\label{fig_fit_CFIRB} CFIRB models induced from the luminosity density
shown in Fig. \ref{fig_sfr}a for the three luminosities 
(continuous line: 3~10$^{12}$ L$_{\odot}$, dotted line: 5~10$^{11}$~L$_{\odot}$, 
dashed line: 7~10$^{10}$~L$_{\odot}$), a cosmology h=0.65, $\Omega_{0}$=1,
$\Omega_{\Lambda}$=0. 
Also shown are the observational constraints
and the CFIRB FIRAS and DIRBE spectrum (stars with error bars) computed as explained in Sect. \ref{CFIRB_base}.}
\end{figure*}

\section{\label{inversion} Luminosity density history from Monte Carlo
simulations}

\subsection{\label{CFIRB_base} The CFIRB spectrum}

We used the CFIRB determination as described in Lagache et al. (2000).
The CFIRB resulting from an integral over a significant
redshift range is expected to be a smooth function of frequency.
Therefore, we use a smooth fit for the determination of $\varphi(z)$.
The CFIRB spectrum, can be fitted between 200 and 2000~$\mic$ by a modified Planck 
function as given by:
\begin{equation}
\label{analy_CFIRB}
I(\nu)=8.80 \times 10^{-5} (\nu / \nu_0)^{1.4} P_{\nu}(13.6 K)
\end{equation}
where $\nu_0$=100~cm$^{-1}$.
The set of parameters (T, $\tau$, $\alpha$) has been determined by
a $\chi^2$ minimization. The spectrum is sampled at 10 FIRAS frequency.
The uncertainties on each sampled frequency are
obtained by varying parameters until $\chi^2$ is increased by 10$\%$.
In addition to the FIRAS data, we have used the CFIRB DIRBE determinations
at 100, 140 and 240 $\mu$m (Lagache et al. 2000). CFIRB data points
and uncertainties that are used for the determination of $\varphi(z)$
are shown in Fig. \ref{fig_fit_CFIRB}.

\subsection{Determination of $\varphi(z)$}

We determine the range of functions $\varphi(z)$ allowed by the data for each combination of SED and 
cosmological model. 
The cases used are given by the combinations of:
\begin{itemize}
\item{Three IR galaxy luminosities (3 10$^{12}$ L$_{\odot}$, 5 10$^{11}$ L$_{\odot}$ and 
7 10$^{10}$ L$_{\odot}$, see Fig. \ref{fig_spec_gal})}
\item{Four values for the dust spectral index (1.3, 1.5, 1.7 and 2)}
\item{Three cosmological
models defined by the set of parameters h, $\Omega_0$ and $\Omega_{\Lambda}$
(h=0.65, $\Omega_0$=0.3, $\Omega_{\Lambda}$=0.7; h=0.65, $\Omega_0$=0.3, $\Omega_{\Lambda}$=0, 
and h=0.65 $\Omega_0$=1, $\Omega_{\Lambda}$=0) which fix dt/dz.}
\end{itemize} To establish $\varphi(z)$ and the associated error bars
we use Eq. (\ref{eq_principale}).
The basic algorithm for finding N$_z$ is based on Monte Carlo simulations.
N$_z$ is sampled at a few redshift values and linearly interpolated between these values
for computing the term on the right side of Eq. (\ref{eq_principale}). 
It has been assumed that beyond z=13, N$_z$=0. For each case, the solution (minimum $\chi^2$) has been obtained 
by exploring a wide range of randomly distributed
values of  N$_z$ at each of the sampled values of z. Error bars are estimated by keeping the computed 
CFIRB within the uncertainties at each sampled frequency (see Fig. \ref{fig_fit_CFIRB}) and greater than 
the lower CFIRB limit at 850 $\mu$m from Barger et al. (1999).
By using an iterative method for progressively reducing the range of explored values, 
we reach convergence for each case studied with a reasonable number of 
hits ($\sim$ 50000). At z=0, the IR production rate in the Universe is around 
1.65 10$^8$ h L$_{\odot}$/Mpc$^{3}$ (Soifer \& Neugeubauer 1991),
where h is the Hubble constant in units of 100 km/sec/Mpc$^{3}$.
This gives $\varphi(z=0)$= 10$^8$ L$_{\odot}$/Mpc$^3$ for h=0.65.

\subsection{\label{sect_results} Results}

The results{\footnote{Figures and results can be found on G. Lagache's WEB page:
http://www.ias.fr/iasnv/people.html}}
are summarised in Table \ref{tbl-1} for the different cosmologies and SEDs. 
We see that for all the considered cases, the $\chi^2$ is very similar. 
This shows that, as expected, there is not a unique solution for the inversion
in terms of cosmological model and average SEDs.
Nevertheless the remarkable result is that there is a range of redshifts in which all acceptable solutions have 
the same behaviour as can be seen in Fig. \ref{phi_all} where the luminosity density variation, 
together with its uncertainties allowed by the data, is shown for each case.
On the one hand, as could be expected, constraints on $\varphi(z)$ are very weak below
redshift 1 (no cosmic background values at mid-IR wavelengths) 
and above redshift 4 (very low signal
to noise ratio of the CFIRB spectrum above 800 $\mu$m). On the other hand between redshifts
1 and 4, the CFIRB gives strong constraints on the history of the far-IR production rate which 
cannot be established from any of the present source surveys. The luminosity density
is about 10 times higher at z=1 than at z=0 and it 
is nearly constant up to redshift 4. Because of this rather constant
behaviour, the level is only weakly dependent on the galaxy
model spectrum taken with different peak wavelength (see Fig. \ref{fig_sfr}) as a change in the peak wavelength 
mainly shifts the function $\varphi(z)$ in z.
Moreover, the luminosity density in the redshift range 1 to 4 is not affected by changing the
far-IR dust spectral index from 1.3 to 2.
On Fig. \ref{fig_fit_CFIRB} is
shown the CFIRB models induced from the luminosity density
variation for the three luminosities and a fixed cosmology.
For the far-IR part of the spectrum, the different luminosities
give automatically good fits . This is not necessarily the case below 100 $\mu$m where
the result is very dependent of the galaxy model spectrum. 
With the present observational constraints, the best fit is obtained
for a luminosity of 5 10$^{11}$ L$_{\odot}$. This is very consistent with
typical luminosities of IR galaxies that are
making the background at 15 $\mu$m (Aussel, 1999).

\begin{figure}  
\epsfxsize=9.cm
\epsfysize=7.cm
\epsfbox{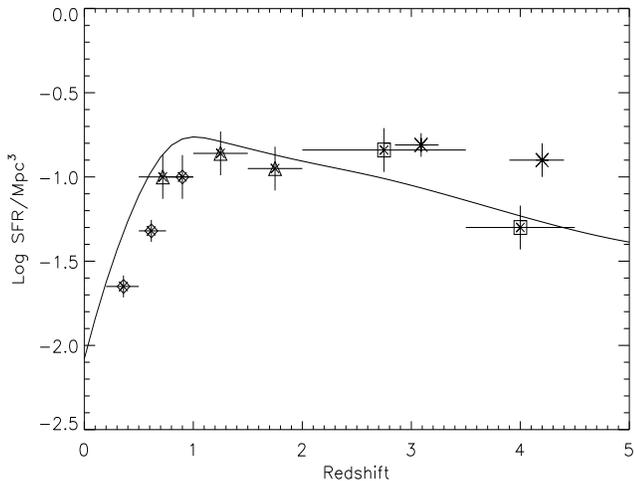}
\caption{\label{fig_sfr_all} SFR derived from UV/Vis/Near-IR
observations (diamonds: Lilly et al. 1996; triangles: Conolly et al. 1997;
squares: Madau et al. 1996 and crosses: Steidel et al. 1999)
corrected for extinction by Steidel et al. (1999) and SFR derived from the CFIRB 
(continuous line, scaled to H$_0$=50 km s$^{-1}$ Mpc$^{-1}$
for consistency)}
\end{figure}

\begin{table}
\caption{SFR (in M$_{\odot}$ yr$^{-1}$ Mpc$^{-3}$)
deduced from the CFIRB and from deep submm SCUBA surveys 
(for H$_0$=65 km s$^{-1}$ Mpc$^{-3}$).} 
\label{tabSCUBA}
\begin{center} 
\begin{tabular}{|l|c|c|} \hline
Publications & Redshift & SFR \\ \hline
This paper & z=1.9 & 0.21$\pm$0.10 \\ 
Barger et al. 1999 & 1$<$z$<$3 & 0.25 \\
Hughes et al. 1998 & 2$<$z$<$4 & $>$0.14 \\ \hline
\end{tabular}\\
\end{center}
\end{table}

\section{\label{sfr} Cosmological implications}

\subsection{The star formation history of the Universe}
Many models describing the evolution of galaxies including their IR and submm
emission have been published in the recent years.  
In this section, we discuss only empirical determinations of the
SFR derived from different observations.\\

The history of the cosmic Star Formation Rate (SFR) can be derived
from deep optical surveys assuming that (1) the stellar Initial Mass
function (IMF) is universal, (2) the far-UV light is proportionnal
to the SFR and (3) extinction is negligible. The presence of dust
which absorbes most of the UV starlight in starburst galaxies
makes this last assumption highly questionable.
The corrections needed to account for extinction
are rather uncertain and there is much controversy about
the value of this correction. Moreover, the SFR deduced from
optical surveys can be underestimated if there is a significant
population of objects so obscured that they are not detected
in these surveys. A direct determination of the fraction of the stellar radiation reradiated by dust can 
be obtained from the IR/submm surveys if the dust enshrouded AGNs do not dominate. However, so far, the catalogues
of faint submm sources with reliable redshifts are not large 
enough to reconstruct the history of the SFR (see Lilly et al. 1999
for a first attempt). We have shown that strong constraints can be provided by the CFIRB in the redshift 
range 1 to 4. It is particularly interesting to compare our determination with (1) the
optically-derived SFR, corrected for extinction in the same redshift range and 
(2) the SFR inferred from SCUBA submm surveys. However, 
it is beyond the scope of this paper to discuss the cosmological
implications in term of star formation. Detail discussions
exist for example in Dwek et al. (1998), Pei et al. (1999), and Madau (1997).\\

Fig. \ref{fig_sfr_all}  shows a compilation of the SFR derived
from UV/Vis/Near-IR surveys, together with our determination
of the far-IR radiation production rate history (through an interpolation). 
To compare the star formation rate with the luminosity density, 
we have to use a conversion factor. 
We take $\frac{SFR}{M_{\odot} yr^{-1}}= \frac{L_{IR}}{7.7 10^9 \hspace{0.07cm} L_{\odot}}$
(Guiderdoni et al. 1998, with a Salpeter IMF), conversion which is in good agreement with that
derived from Scoville \& Young (1983) and Thronson \& Telesco (1986).
The UV luminosity density in Fig. \ref{fig_sfr_all} has been corrected for extinction by
Steidel et al. (1999).
We see a very good agreement between the UV and far-IR
luminosity density confirming the need for a very large extinction
correction. 
This is a strong indication that the population of galaxies that are making the submm EB
(300-800 $\mu$m) seems to be the same as the population detected by Steidel et al. (1999) in their surveys 
of Lyman-break galaxies. This would also imply that the population of objects so obscured that they are not detected
in UV/opt/near-IR surveys cannot contribute for a large fraction of
the luminosity density. \\

Submm EB deduced SFR can also be compared to SCUBA results.
Several groups are now conducting deep and ultradeep blanck-field
surveys (Barger et al. 1999; Hughes et al. 1998; Eales et al. 1999), 
that follow the first survey of Smail et al. (1997)
who discover a population of luminous galaxies emitting at 850 $\mu$m
amplified by lensing from foreground clusters.
Two of these groups (Hughes et al. 1998; Barger et al. 1999)
currently give estimates of the submm source SFR whereas Lilly et al.
(1999) discuss its probable behaviour. Table \ref{tabSCUBA}
compares the SFR derived from the submm
EB and the SCUBA determination. There is a very good agreement
between the two although the SCUBA estimates should be interpreted
with caution since only 20-25 $\%$ of the detected sources have 
secure identifications (Sanders, 1999).\\
Fig. \ref{fig_fit_CFIRB} shows that all galaxies that are contributing
to most of the background at 850 $\mu$m cannot have a luminosity
greater than 2-3 10$^{12}$ L${\odot}$, standard luminosities of the
current detected SCUBA sources (Lilly et al. 1999; Eales et al. 1999).
This is consistent with the fact that SCUBA sources above 3 mJy account for
only 20-30 $\%$ of the EB at 850 $\mu$m. The present data show that the bulk
of the submm EB is likely to reside in sources with 850 $\mu$m fluxes near 1 mJy.
Barger et al. (1999) estimate that the FIR luminosity of a characteristic
1 mJy source is in the range 4-5 10$^{11}$ L$_{\odot}$, which is what is expected from Fig. 
\ref{fig_fit_CFIRB}.\\

It can be checked that models of galaxy evolution which fit the CFIRB fall
within the allowed range of SFR histories obtained here (see for example Pei et al., 1999 and
Guiderdoni et al., 1998).

\subsection{Redshift contribution}

\begin{figure}  
\epsfxsize=9.cm
\epsfysize=7.cm
\epsfbox{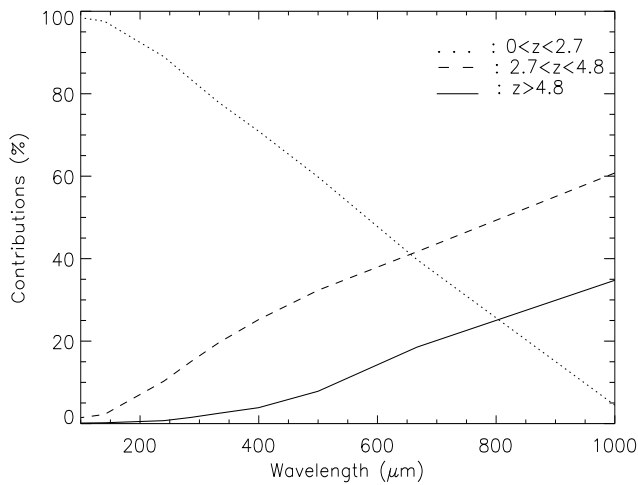}
\caption{\label{fig_contrib} Relative contribution to the CFIRB of galaxies
with luminosity of 5~10$^{11}$~L$_{\odot}$ for z$<$2.7
(dotted line), 2.7$<$z$<$4.8 (dashed line), and z$>$4.8
(continuous line).}
\end{figure}

On Fig. \ref{fig_contrib} is shown the relative contribution from different redshift ranges to the
CFIRB as a function of wavelengths 
(for one illustrative case $\alpha$=2, L=5 10$^{11}$ L$_{\odot}$, $\Omega_0$=0.3 
and $\Omega_{\lambda}$=0.7). 
As expected, galaxies below redshift 2.7 contribute
mostly at short wavelengths and galaxies above redshift
2.7 contribute mostly at long wavelength. 
As can be seen in Fig. \ref{fig_sfr}, the small error bar
at z$\sim$2.5 shows that this conclusion is firmly
established. On the contrary, the fraction shown in Fig. \ref{fig_contrib} 
for redshift larger than 4.8 is not
well constrained by the background.

At 850 $\mu$m, about 20$\%$ of the CFIRB light
comes from galaxies below redshift 2.7 and 80$\%$ from
above. 
This can be compared with the recent results
of Eales et al. (1999). Their data suggest that at least 15$\%$
of the 850 $\mu$m EB is emitted at z$<$3, which
is in good agreement with what we obtain here.
However, our results show that a significant
fraction of the submm EB comes from very
distant objects. Such a scenario represents a picture in which 
a significant fraction of all stars has been formed
very early in the Universe.

\section{Summary}
The CFIRB detected in the COBE data at wavelength greater than 100 $\mu$m contains a 
surprisingly large fraction of the cosmic background due to distant galaxies. 
The spectrum of this background at long wavelength is significantly flatter than the one observed 
for individual IR galaxies. This implies that the submm part of the CFIRB cannot be dominated 
by the emission of the galaxies which account for most of the CFIRB at 150 $\mu$m, and thus contains a unique 
information about high redshift IR galaxies.
Considering the variety of long wavelength spectra observed for these 
galaxies we have explored the range of possible 
redshift evolution histories. We show that only a
co-moving production rate of far-IR 
radiation with strong evolution at low redshifts but little evolution 
between redshifts 1 and 4 is the only solution allowed by the CFIRB (the 
detailed low redshift evolution is much better constrained
by the ISO deep surveys than by the background). 
Our results show that there is no evidence for a ``peak'' in the cosmological 
star formation density
at z=1-2 as it has been assumed by many authors; it is clear that the
epoch of the beginning of star formation has not yet been identified.
Moreover, these results indicate that there is a divergence 
of the behaviour of the star formation history as compared to that of the space
density of luminous AGNs (see for example Dunlop, 1997 and 
Shaver et al. 1998), as it was aslo suggested by Steidel et al. (1999).

\acknowledgements
We thanks Martin Harwit for very useful discussions on the content
of this paper. The work presented here profited from very
usefull comments from the referee.\\

{\bf Appendix 1: Observational constraints on the cosmic background}
\par\bigskip

At wavelengths at which the contribution is not negligible the CMB 
as well as its dipole distorsion have been removed using the values
given in Mather et al. (1994) and Fixsen et al. (1994).
In the following, cosmic background is used to refer to the extragalactic
electromagnetic radiation background outside the CMB.\\

{\bf Far-IR and submillimeter range ($\lambda\ge100 \mic$):}\\

In this wavelength domain, the determination of the cosmic background
has been done using the COBE FIRAS and DIRBE data. The main difficulty
is to remove the bright and fluctuating Galactic dust emission.
The first detection has been reported by Puget et al. (1996).
They used an independent dataset (the HI 21cm survey
of Hartmann et al. 1994) in addition to FIRAS spectra
to remove the dust emission. The detection of the background
at $\lambda\ge140 \mic$ has been later confirmed by Fixsen et al. (1998), 
Hauser et al. (1998), Schlegel et al. (1998), Lagache et al. (1999) and Lagache et al. (2000), 
using the DIRBE and FIRAS data. 
All determinations are today in good agreement for $\lambda>$150~$\mu$m.
Discrepancies at 140 $\mu$m are due to different evaluations of
the emission from dust associated with the ionised gas in the interstellar
medium{\footnote{The CFIRB value of Lagache et al. (1999) is smaller 
than that derived in Lagache et al. (2000) since the assumed WIM (Warm Ionised
gas) dust temperature
was overestimated (the WIM dust spectrum was
very noisy below 200~$\mu$m and the estimated dust
temperature was too high).}}.
At 100~$\mu$m, 
assuming an accurate subtraction of the zodiacal emission,
Lagache et al. (2000) using 
two independent gas tracers for the HI (Leiden-Dwingeloo survey)
and the H$^+$ (WHAM data) obtain
I$_{CFIRB}$(100)= 0.78$\pm$0.21 MJy/sr. One has to note that methods based on
the intercept of the far-IR/HI correlation
for the determination of the CFIRB 
are dangerous. For example, for the parts of the sky covered
by WHAM, this intercept is about 0.91 MJy/sr, which is significantly different from
the value of the CFIRB (0.78 MJy/sr). It is interesting to compare
the CFIRB value of 0.78 MJy/sr to the non-isotropic residual
emission found by Hauser et al. (1998). The average over three regions
of the residual emission, equal
to 0.73$\pm$0.20 MJy/sr, is in very good agreement with the
Lagache et al. (2000) determination. \\

{\bf Mid IR range ($10 \le \lambda \le 80 \mic$):}\\

In this wavelength domain, the level of the interplanetary dust emission
is very high with respect to the expected level of the cosmic background.
This interplanetary dust emission peaks around 25$\mic$.
Around this wavelength, since the level of such an emission
is around a thousand times the level of the background,
a direct determination of the background from an earth orbit is impossible.
At 25 and 60 $\mic$, our description of the interplanetary
dust emission is not accurate enough to isolate
the background emission. Therefore, in the Mid IR
range, only indirect determinations of the background are possible. Lower
limits are provided by deep cosmological surveys.
We have reported on Fig. \ref{back_all} lower limits
derived from surveys carried out with the ISOCAM instrument aboard the IR Space Observatory:
at 6.5~$\mic$ (D\'esert, private communication), 12 $\mic$ 
(Clements et al. 1999) and 15 $\mic$ (Elbaz et al. 1999). 
At 15 $\mic$, two values are reported: a lower limit and an absolute value. 
The lower limit is derived
from the integrated brightness for sources above
50~$\mu$Jy. As the counts flattens significantly below
100~$\mu$Jy it is interesting to display 
the integrated brightness down to very low fluxes using
an extrapolation of the counts with a model fitting the data
(Guiderdoni et al. 1998), as the result does not depend
much on the model used. We have also reported a lower
limit at 60 $\mic$ from IRAS galaxy counts (Lonsdale et al. 1990).\\
 Upper limits are provided by the
propagation of very high energy gamma rays (in the TeV range) through 
intergalactic space. The principle of such a
determination is the following: gamma rays are absorbed by pair creation
on the photons of the cosmic background. For a gamma ray of energy E coming from 
a distant extragalactic source all cosmic background photons with energy above a 
threshold $\epsilon=\frac{2 m_e c^{2}}{E}$ will contribute to the absorption of 
the gamma ray. As more photons contribute to
the absorption when the gamma ray energy increases the gamma ray spectrum
should present a curvature at high energy if the integrated density of photons is 
high enough. Early determinations (Stecker \& De Jager, 1993; 
Biller et al. 1995) relied on assumptions 
on the shape of the spectrum at the source (a power law extrapolated from lower 
energies). Recently variations of the spectrum of the source Mkr 501 have been 
detected with CAT in a burst mode during which the X-ray emission has been observed.
This allows the prediction, with minimal hypothesis, of the initial source
spectrum and thus get a firm
determination of an upper limit (Barrau 1998; Barrau et al, in preparation). 
This limit is in very good agreement with that of Funk et al. (1998) from HEGRA data.
This leads to upper limits only a factor of 2 above the lower limit given
by the extrapolated source counts. \\

{\bf Ultra-violet, visible and near-IR bands:}\\

Pozzetti et al. (1998) have compiled number counts of extragalactic sources in
the Hubble Deep Field (HDF) in the U, B, V, I and K bands. The integrated background of 
these counts have converged well and thus we already have a good determination of 
the background due to extragalactic sources. 
Part of the stellar radiation has been absorbed by dust and the
corresponding energy has been observed in form of the IR cosmic background. 
Another part  may have been scattered for
example by intergalactic hydrogen clouds into a diffuse background.
Scattering cross sections (by atoms, molecules or dust) decrease fast with 
wavelength thus any scattered isotropic component should be detectable primarily 
at ultraviolet wavelengths. Direct measurements of the
extragalactic background are very difficult in the visible range because at
these wavelengths, the scattered solar radiation by interplanetary particles is 
larger than the expected cosmic background by a factor one hundred. Fortunately 
in the ultraviolet the fast decrease of the solar spectrum allows a good determination 
of an upper limit (Bowyer 1991; Martin et al. 1991). This limit at 1500{\AA}
is shown in Fig. \ref{back_all} and is at comparable level as the lower limit from
the HDF counts in U band (at 3600{\AA}). Also reported in Fig. \ref{back_all}
is the determination of Armand et al. (1994) which is in line of
the other determinations. The Armand et al. value is based
on the extrapolation of galaxy counts at 2000{\AA} based on the balloon-borned
experiment FOCA (Milliard et al. 1992).
We thus conclude that the scattered part 
cannot exceed 50~$\%$  in the ultraviolet and is very likely to be negligible in 
the B, V, I and K bands. We can 
thus make the conservative hypothesis that the HDF counts provide measurements of the
cosmic background. However, recent direct measurements of the optical
background at 3000, 5500 and 8000 {\AA} from absolute surface
photometry by Bernstein et al. (in preparation, quoted by Madau \&
Pozzetti, 2000) lie between a factor of 2.5 to 3 higher than the
integrated light from galaxy counts, with an uncertainty
that is largely due to systematic rather than statistical errors. This direct determination
is very difficult because of the uncertainties level of the zodiacal background
to be subtracted.\\
Also reported
in Fig. \ref{back_all} are the direct estimation of the background of
Dwek \& Arendt (1998) with the DIRBE 3.5 $\mic$ band and 
Gorjian et al. (2000) at 2.2 and 3.5 $\mu$m. This determination is very 
difficult due to the contamination by the emission of stars. \\
The upper limit from high energy gamma
rays at 10 $\mu$m from CAT combined with the extrapolation
of the integrated flux from the 15 $\mu$m counts, and
the decrease of the integrated emission from sources in the ultra deep
ISOCAM surveys between 15 and 6 $\mu$m,  give
an indication that the background spectrum must present a minimum
between 3 and 10 $\mu$m. This minimum is expected to be the wavelength
at which the direct stellar contribution becomes equal
to the dust (in fact PAHs) emission.\\

\par\bigskip
{\bf Appendix 2: Far-IR radiation production rate for a single frequency emitting galaxy}
\par\bigskip
If one makes the extreme assumption that the radiation of IR galaxies 
is concentrated at a single frequency, the spectrum of the background can 
be easily inverted to get the radiation production rate.
If the production rate of IR radiation per co-moving volume element
at a single frequency
$\nu_o$ is $\phi_{ \nu_o}(z)$, the resulting cosmic background in the 
simple case of Euclidean Universe
will be given, following  Eq. 2, by:
$$
{\phi_{ \nu_o}( z) \over L_{\odot} Mpc^{-3}}
 = 6.5\, 10^{7} {\nu I_{\nu} \over 10^{-8} Wm^{-2}sr^{-1}}
(\nu / \nu_o)^{-5/2} 
$$

For a simple approximation of the background with power laws 
$$
\nu I_{\nu} = 12\, 10^{-8}(\nu / \nu_o)^{2.5}  Wm^{-2}sr^{-1}
$$
valid 
between 200 $\mu$m and 500 $\mu$m, and
$$
\nu I_{\nu} = 27\, 10^{-8}(\nu / \nu_o)^{3}  Wm^{-2}sr^{-1}
$$
valid 
between 500 $\mu$m and 1.5 mm,
($\nu_o= 3 10^{12} $Hz),
we deduce
$$
\phi_{ \nu_o}( z) = 7.8\, 10^{8}  L_{\odot} Mpc^{-3}
$$
for  $1 < z < 4$
and
$$
\phi_{ \nu_o}( z) = 1.7\, 10^{9} (1+z)^{-0.5} \left( {\nu_o \over {3 10^{12} Hz}} \right)^{0.5} 
L_{\odot} Mpc^{-3}
$$
for $z > 4$.\\
There is {\it no dependance} of the energy production rate with $\nu_o $ for 
z between 1 and 4 because the slope of the background corresponds to the 
case of no evolution of the co-moving radiation production rate.
This illustrates why $\varphi(z)$ does not depend very much on the spectrum of
the galaxy model in this redshift range.\\


\end{document}